\documentclass{emulateapj}
\newcommand{\km}{${\rm km\,s}^{-1}$}
\newcommand{\fuse}{{\em FUSE}}
\newcommand{\hi}{H$\;${\small\rm I}\relax}
\newcommand{\hii}{H$\;${\small\rm II}\relax}

\newcommand{\cii}{C$\;${\small\rm II}\relax}

\newcommand{\ciii}{C$\;${\small\rm III}\relax}
\newcommand{\civ}{C$\;${\small\rm IV}\relax}
\newcommand{\nni}{N$\;${\small\rm I}\relax}
\newcommand{\nii}{N$\;${\small\rm II}\relax}

\newcommand{\nv}{N$\;${\small\rm V}\relax}
\newcommand{\oi}{O$\;${\small\rm I}\relax}
\newcommand{\oii}{O$\;${\small\rm II}\relax}
\newcommand{\oiii}{O$\;${\small\rm III}\relax}

\newcommand{\ovi}{O$\;${\small\rm VI}\relax}

\newcommand{\mgii}{Mg$\;${\small\rm II}\relax}
\newcommand{\sii}{S$\;${\small\rm II}\relax}
\newcommand{\siii}{Si$\;${\small\rm II}\relax}
\newcommand{\siiii}{Si$\;${\small\rm III}\relax}
\newcommand{\aliii}{Al$\;${\small\rm III}\relax}
\newcommand{\Siii}{S$\;${\small\rm III}\relax}
\newcommand{\siiv}{Si$\;${\small\rm IV}\relax}

\newcommand{\svi}{S$\;${\small\rm VI}\relax}
\newcommand{\feii}{Fe$\;${\small\rm II}\relax}
\newcommand{\feiii}{Fe$\;${\small\rm III}\relax}

\newcommand{\lya}{Ly\,$\alpha$\relax}
\newcommand{\lyb}{Ly\,$\beta$\relax}
\newcommand{\lyg}{Ly\,$\gamma$\relax}

\slugcomment{Accepted for Publication in the ApJ}
\shortauthors{Lehner et al.}
\shorttitle{QSO Asorbers-Galaxies Connection}
\begin{document}

\title{The Connection between a Lyman Limit System, a very strong \ovi\ Absorber, and Galaxies at $z\sim 0.203$\altaffilmark{1} }
\author{N.\ Lehner\altaffilmark{2},
	J.X. \ Prochaska\altaffilmark{3},
	H.A. \ Kobulnicky\altaffilmark{4},
	K.L. \ Cooksey\altaffilmark{3},
	J.C. \ Howk\altaffilmark{2}, 
	G.M. \ Williger\altaffilmark{5},
	S.L. \ Cales\altaffilmark{4}
	}
   
\altaffiltext{1}{Based on observations made with the NASA-CNES-CSA 
Far Ultraviolet Spectroscopic Explorer. FUSE is operated for NASA by the Johns 
Hopkins University under NASA contract NAS5-32985. Based on observations made with the NASA/ESA Hubble Space Telescope,
obtained at the Space Telescope Science Institute, which is operated by the
Association of Universities for Research in Astronomy, Inc. under NASA
contract No. NAS5-26555.}
\altaffiltext{2}{Department of Physics, University of Notre Dame, 225 Nieuwland Science Hall, Notre Dame, IN 46556}
\altaffiltext{3}{UCO/Lick Observatory, University of California, Santa Cruz, CA }
\altaffiltext{4}{Department of Physics \& Astronomy, University of Wyoming, 1000 E. University, Laramie, WY 82071}
\altaffiltext{5}{Department of Physics, University of Louisville, Louisville, KY 40292}

\begin{abstract}
With a column density $\log N($\ovi$) = 14.95 \pm 0.05$, the \ovi\ absorber at $z_{\rm abs} \simeq 0.2028$ 
observed toward the QSO PKS\,0312--77 ($z_{\rm em} = 0.223$) is the strongest yet detected 
at $z < 0.5$. At nearly identical redshift  ($z_{\rm abs} \simeq 0.2026$),  we also identify a 
Lyman limit system (LLS, $\log N($\hi$) = 18.22\,^{+0.19}_{-0.25}$). 
Combining FUV and NUV spectra of PKS\,0312--77 with optical observations of
galaxies in the surrounding field ($15\arcmin \times 32\arcmin$), we present an analysis of these absorbers 
and their connection to galaxies. The observed \oi/\hi\ ratio and photoionization modelling of other low ions indicate
the metallicity of the LLS is $[{\rm Z/H}]_{\rm LLS} \simeq -0.6$ and that the LLS 
is nearly 100\% photoionized. In contrast, the \ovi-bearing 
gas is collisionally ionized at $T\sim (3$--$10) \times 10^5$ K as derived from the high-ion ratios
and profile broadenings. Our galaxy survey 
reveals 13 ($0.3 \la L/L_* \la 1.6$) galaxies at $\rho < 2 h^{-1}_{70}$ Mpc
and $|\delta v| \la 1100$ \km\ from the LLS. A probable origin for 
the LLS is debris from a galaxy merger, which led to a  $0.7 L_*$ galaxy ($[{\rm Z/H}]_{\rm gal} \simeq +0.15$) at
$\rho \simeq 38 h^{-1}_{70}$ kpc. Outflow from this galaxy may  also be responsible for 
the supersolar ($[{\rm Z/H}]_{\rm abs} \simeq +0.15$), fully ionized absorber at $z_{\rm abs} \simeq 0.2018$ ($-190$ \km\ from the LLS).
The hot \ovi\ absorber likely probes coronal gas about the $0.7 L_*$ galaxy and/or ($\sim$0.1 keV) intragroup gas of a spiral-rich system. 
The association of other strong \ovi\ absorbers with LLS suggests they trace galactic and not intergalactic structures.
\end{abstract}

\section{Introduction}
Connecting the QSO absorbers with their environments is crucial for using the absorbers 
to study the properties and evolution of galaxies, the intergalactic medium (IGM), and 
the galaxy-IGM interface.  The taxonomy of QSO absorbers is usually made according 
to their  \hi\ column densities ($N($\hi$)$): the Ly$\alpha$ forest \citep{rauch98}
with $\log N($\hi$) < 16$, the Lyman limit systems that are optically thick at the Lyman limit 
\citep[LLS;][]{tytler82} with $16 \le \log N($\hi$) < 20.3$, and the damped Ly$\alpha$ absorbers 
\citep[DLAs;][]{wolfe05} with $\log N($\hi$)  \ge 20.3$. 
Given their  placement in the \hi\ column density hierarchy, the LLS likely represent the
interface  between the tenuous, highly ionized Ly$\alpha$ forest  and the dense, neutral DLAs. 

Many observational studies have been undertaken to connect these absorbers to physical objects,
such as galaxies, galaxy halos, intergalactic voids. At $z\la 1$, analyses of the  absorber-galaxy relationship 
show that QSO absorbers are not distributed randomly with respect to galaxies, even 
for absorbers with the lowest \hi\ column densities detected so far 
\citep[e.g.,][]{lanzetta95,tripp98,impey99,chen01,bowen02,penton02,stocke06,wakker08}. 
Strong  \mgii\ absorbers are  found within the extended halos (impact parameter $\rho < 100$ kpc) 
of individual galaxies \citep[e.g.,][]{bergeron91,steidel93}, while the DLA are generally associated with the 
main bodies of galaxies \citep[e.g.,][]{chen03,rao03,wolfe05}. Although strong \mgii\ absorbers are believed
to be tracers of LLS, there is a large scatter between $W_\lambda($\mgii$)$ and
$N($\hi$)$, not allowing a direct connection between these two quantities \citep{churchill05,menard08}, but see
also \citet{bouche08}.
Furthermore the origin, metallicity, and physical properties of the \mgii-bearing gas is 
largely unknown because of the limited information available (e.g., metallicity, ionization and physical
conditions).  This requires high spectral resolution
UV space-based observations at $z\la 1$ where several \hi\ Lyman series lines 
and metal lines in various ionization stages combined with galaxy redshift surveys and imaging 
can be acquired. 

Currently, there are only three reported LLS where detailed information 
about their metallicity, ionization, and galaxy environment is available. 
Those are at $z = 0.08092 $ toward PHL\,1811 \citep{jenkins03,jenkins05}, 
$z =  0.16710$ toward PKS\,0405--123 \citep{chen00,prochaska04,prochaska06,williger06}, 
and $z = 0.09847 $ toward PKS\,1302--102  \citep{cooksey08}. 
These studies suggest that LLS can be either  metal-rich ($Z$$\sim$$0.5$--$1 Z_\odot$) or metal poor 
($Z\la 0.02 Z_\odot$) and are generally associated with the extended  reaches ($\sim30$--100 kpc) of individual galaxies.
Combined with previous studies, these conclusions demonstrate that LLS are crucial for 
understanding the interaction between galaxies and their environments, an important ingredient
in any cosmological simulations \citep[e.g.,][]{bertone07,oppenheimer08a}.

In this paper, we present another detailed study of the relationship between a LLS observed 
at $z= 0.20258$ toward the QSO PKS\,0312--77 ($z_{\rm em} = 0.2230$) and 
galaxies.\footnote{We also refer the reader to the master's thesis of Sarah S. Giandoni (New Mexico
State University, 2005) for a complimentary analysis of the PKS0312--77 sightline.} 
This QSO was originally observed in the UV using the high-resolution mode of Space Telescope Imaging
Spectrograph (E140M, E230M) onboard of the {\em Hubble Space Telescope}\ ({\em HST}) to study the Magellanic 
Bridge  \citep[see][]{lehner08}, a region of gas linking the Small and Large Magellanic 
Clouds (SMC, LMC). The STIS spectrum revealed very strong \lya\ and \lyb\ absorption lines, 
suggesting the presence of a LLS. This  was subsequently confirmed
by {\em Far Ultraviolet Spectroscopic Explorer}\ ({\em FUSE}) observations where 
the expected flux decrement at the Lyman limit  at $z\approx 0.203$ is detected. We show that the
\hi\ column density (that often lacks in low resolution surveys of Lyman thick absorbers) can be 
determined owing to the high spectral resolution of these spectrographs. Determining 
$N($\hi$)$ is crucial as it allows us to determine the metallicity and ionization conditions of the LLS.

At a similar redshift, this sightline also reveals the strongest \ovi\ absorber  
discovered to date, with $\log N($\ovi$) = 14.95 \pm 0.05$ (see \S\ref{ssec-metal}). 
Strong \ovi\ absorbers are rare and are unlikely to be related to the 
warm-hot intergalactic medium \citep[WHIM,][]{cen99,dave99}, where a large fraction 
of the baryons could reside at $z<0.5$. \citet{oppenheimer08} argued that strong \ovi\
collisionally ionized absorbers are related to the recycling of gas between the 
IGM and galaxies. Their conclusions are based on {\sc Gadget-2} cosmological simulations
that included a variety of wind models and input physics variations. 
The strong \ovi\ absorbers may therefore be higher redshift counterparts of \ovi\ absorption that probes 
the extended halos of galaxies, such as in our Galaxy 
\citep{savage03} and the LMC \citep{howk02,lehner07}. The strong \ovi\  absorbers may 
also probe cooling gas from intragroup medium revealed by soft X-rays observations 
or even possibly relatively cool (0.1--0.3 keV) intragroup gas \citep{mulchaey96}. 
Recent surveys at low $z$ show indeed that, as for the \hi\ and low-ion absorbers, \ovi\ absorbers
are not usually found in intergalactic voids but within $\la 600$--800 kpc of 
galaxies \citep{stocke06,wakker08}. These works, however did not address explicitly the 
origin(s) of very strong ($\log N >14.5$)  \ovi\ absorbers. Hence this sightline provides the 
unique opportunity to study simultaneously the interaction of galaxies with their surroundings using 
very different tracers of gas-phases and energies and to test recent cosmological simulations.

The organization of this paper is as follows. After describing 
the observations and data reduction of the  PKS\,0312--77 UV spectrum in \S\ref{sec-obs}, we 
present our analysis to estimate the redshifts, equivalent-widths ($W_\lambda$), and column densities ($N$),
of the absorbers in \S\ref{sec-anal}. In \S\ref{sec-abs} we determine
the physical properties and abundances of the metal-line absorbers
observed at $z \approx 0.203$, while in \S\ref{sec-gal} we present our galaxies survey
and discuss the relationship between the absorbers and galaxies. In \S\ref{sec-bigpic} we briefly 
discuss that current observations of the galaxy-IGM absorbers show compelling evidence that the LLS and strong \ovi\ absorbers 
trace the galaxy-intergalactic interface, i.e. the galactic environments on physical scale of tens to hundreds of kpc around galaxies. 
A summary of the main results is presented in \S\ref{sec-sum}. 
For the reader's information, all distances in this paper are physical separations derived from the angular diameter
spaces, assuming a $\Lambda$CDM cosmology with $\Omega_m = 0.3$, $\Lambda = 0.7$, 
and $H_0 = 70 $ \km\,Mpc$^{-1}$ (we use the notation $h_{70} = H_0/70\, {\rm km\,s^{-1}\,Mpc^{-1}}$).

\begin{figure}[tbp]
\epsscale{1} 
\plotone{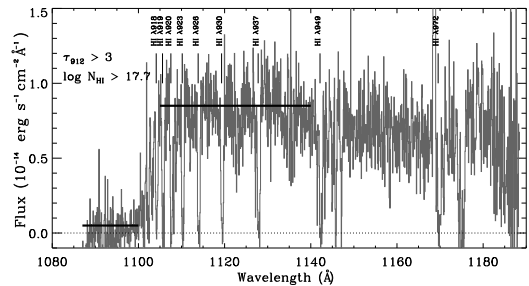}
\caption{Lyman limit at  $z = 0.20258$ observed in the \fuse\ spectrum. The \hi\ lines
used in our profile fitting are indicated (the Ly$\gamma$ line is shown but not used 
because it is contaminated by an airglow emission line). Other absorption lines are either interstellar
or other intervening IGM features.  From the flux decrement, we can place a limit on the 
\hi\ optical depth at the Lyman limit and hence on the \hi\ column density (the flux levels 
to estimate the the optical depth at the Lyman limit are shown with horizontal thick solid lines). For display purposes, 
the data are binned to the {\fuse}\ resolution, i.e. 1 pixel is about 20 \km. 
\label{fig-lls}}
\end{figure}

\section{UV Spectroscopic Observations}\label{sec-obs}
\begin{figure*}[!t]
\epsscale{1} 
\plotone{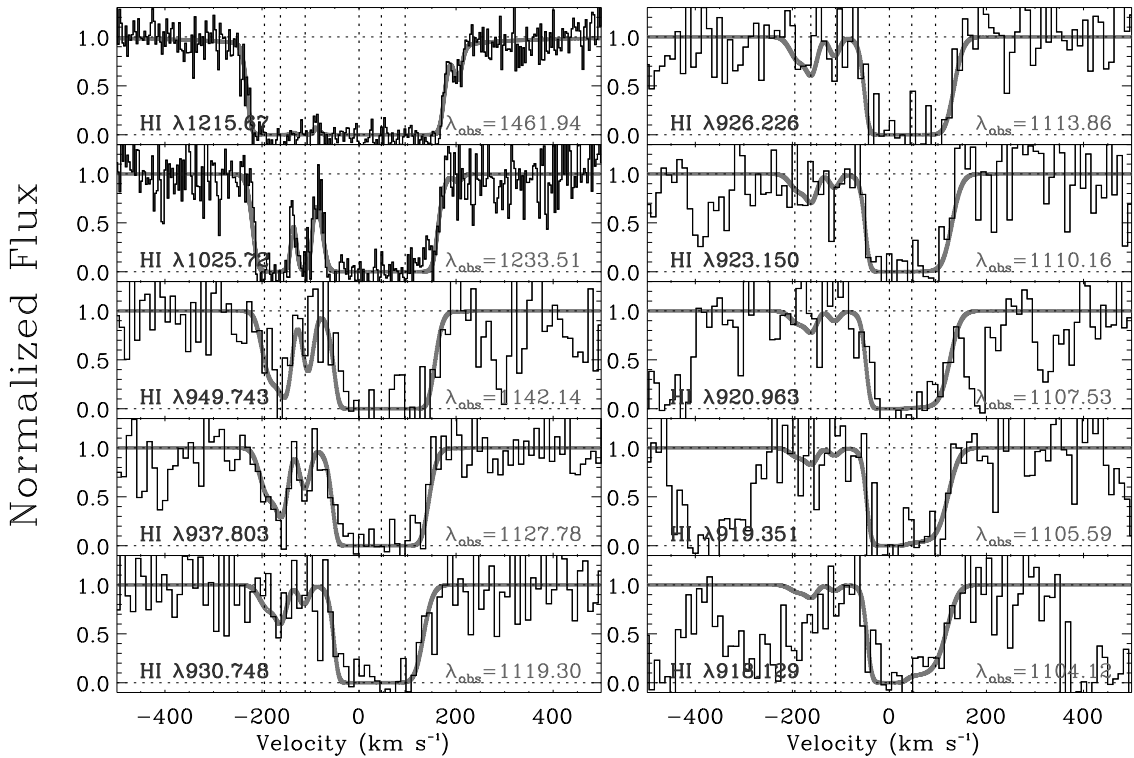}
\caption{Normalized \hi\ profiles relative to $z = 0.20258$.   
The solid line shows our fit to the data.  The vertical dotted lines show 
the velocity centroids of the various components that were used as input in the profile fitting (see Table~\ref{t-hi}). 
\label{fig-hi}}
\end{figure*}

In this work, we present UV observations of the $z\approx 0.203$ absorber detected in 
the  spectra of the QSO PKS\,0312--77. 
The far to near UV spectra of PKS\,0312--77 were obtained 
with \fuse\ (1080--1190 \AA, $R \approx 15,000$; program F108 and E848 -- PI: Lehner and Sembach, respectively) 
and {\em HST}/STIS 140M (1170--1730 \AA, $R \approx 44,000$) and E230M (2130--2980 \AA, $R\approx 30,000$) (program 8651, PI: Kobulnicky). 
The total exposure times are 37.9 ks for the E140M grating, 6.1 ks for the E230M grating,
and 55.5 ks for \fuse\ (for LiF\,2A+LiF\,1B segments). The entrance slit for STIS 
was set to $0\farcs2\times 0\farcs2$, while the $30 \arcsec \times 30\arcsec$ square (LWRS) 
aperture was used for \fuse. Typical signal-to-noise (S/N) 
ratios are about 4--7 per resolution elements (about 7 \km\ for E140M, 10 \km\ for E230M,
and 20 \km\ for \fuse). Unfortunately, only about 20\% of the requested time was obtained
for the \fuse\ program F018 before the failure of the telescope. Yet as illustrated in 
Figs.~\ref{fig-lls} and \ref{fig-hi} the \fuse\ data are particularly useful for
identifying the flux decrement at the Lyman limit at $z=0.20258$ and the Lyman series lines,
which allow a secure determination of the \hi\ column in the $z\approx 0.203$ absorber (see below).  
The full STIS spectrum is shown in  T. Miwava et al. (in prep.), but see 
Fig.~\ref{fig-norm} for the normalized profiles of selected species (see below). 
In the E140M spectrum, the S/N ratio can increase up to about 15
in the QSO emission lines (e.g., near the \nv\ doublet at $z = 0.20258$, S/N\,$\simeq$15). 
On the other hand, at $\lambda \la 1185$ \AA\ and $\lambda \ga 1650$ \AA, the S/N ratio in the
E140M spectrum is very low (S/N\,$\la 2$--3).

The STIS data were reduced using the STIS ID Team version of {\sc calstis} at the 
Goddard Space Flight Center. The procedure for reduction of the STIS E140M data is 
fully described in \citet{tripp08} and includes the two-dimensional echelle scattered light 
correction \citep{valenti02} and the algorithm for automatic repair of hot pixels 
\citep{lindler03}. The {\fuse}\ data were calibrated using the {\sc calfuse} version 
\citep[v3.2,][]{dixon07}. In order to achieve the optimum signal-to-noise, 
LiF\,2A and LiF\,1B segments were coadded.  We note that the continuum of the QSO is not well
behaved on large wavelength scales, but its ``bumpiness" is unlikely due to the ``worm" effect
\citep[see][]{dixon07} since the fluxes in LiF\,1B and LiF\,2A match each other quite well (a similar
behavior is also at longer wavelengths in the STIS bandpass). The fluxes in the individual exposures 
are too low to resolve the interstellar lines, so the separate exposures 
were simply co-added. Inspection of the resulting spectrum and comparison
with common interstellar lines in the STIS spectrum indicates a posteriori
no large shifts were necessary.  
The oversampled {\fuse}\ spectra were binned to a bin size of 0.04 \AA, 
providing about two samples per $\sim$20 \km\ resolution element. 

The profiles were normalized using Legendre polynomials
within $\pm$500--2000 \km\ of each absorption line of interest. The continua 
are simple enough  that the orders of the polynomial used to fit them 
were low ($\le 4$).  

\section{Absorbers Analysis}\label{sec-anal}

To derive the column densities and measure the redshifts, we used the atomic parameters compiled by
\citet{morton03}. 

\subsection{Kinematics Overview}\label{ssec-kin}
\begin{figure*}[tbp]
\epsscale{1} 
\plotone{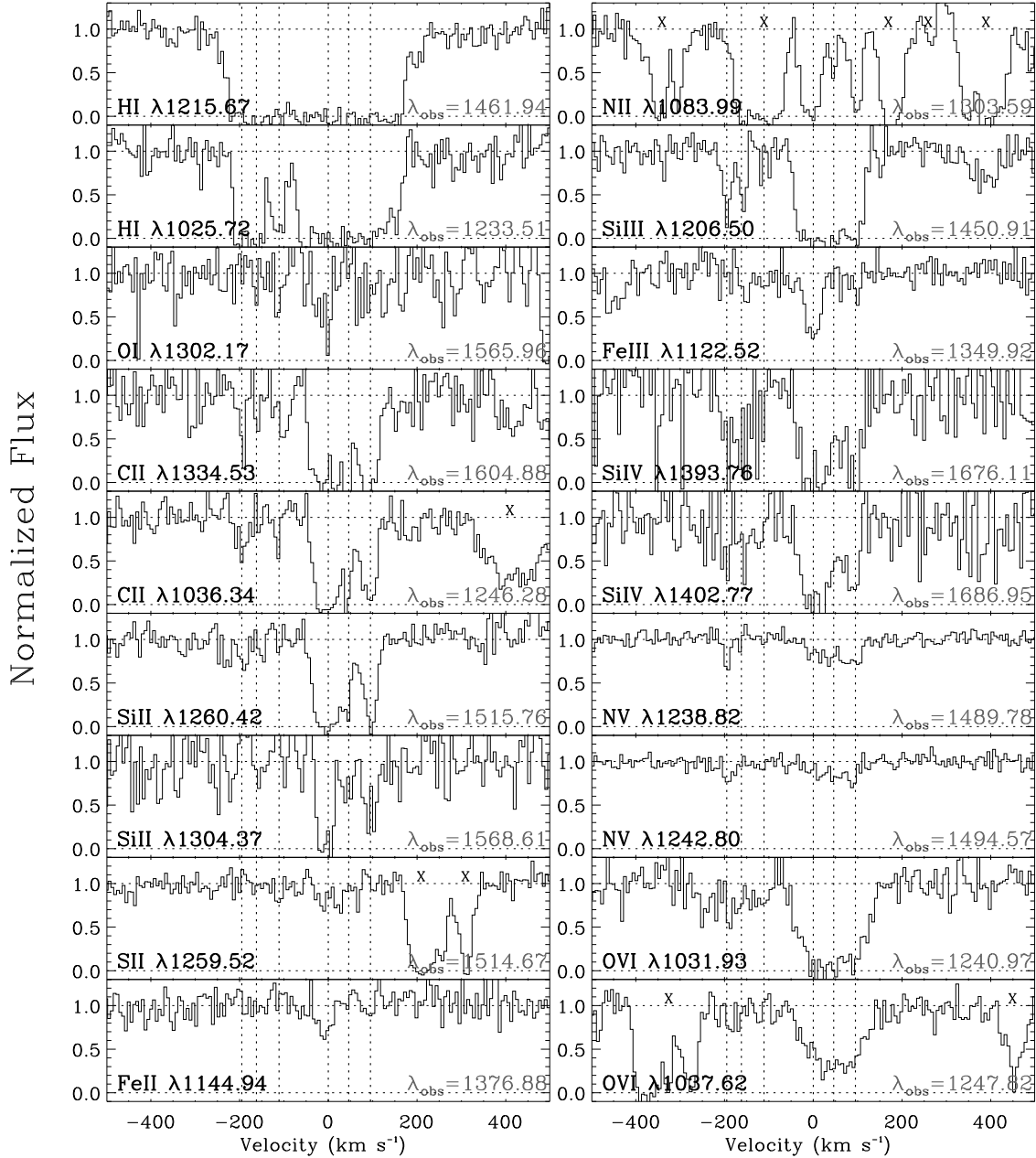}
\caption{Normalized profiles relative to $z = 0.20258$ (ordered approximately by ionization potential).   
The vertical dotted lines show 
the velocity centroids of the various components. The absorber at $z \approx 0.2018$ is 
between $-220$ and $-150$ \km\ in this representation, while the main absorption of LLS 
is between $-40$ and $+30$ \km, but with additional components up to $+130$ \km. 
The ``x'' shows part of the spectrum that is contaminated by other absorbing features. 
\label{fig-norm}}
\end{figure*}

The normalized profiles of the \hi\ and other species are shown in 
Figs.~\ref{fig-hi} and \ref{fig-norm} against the restframe velocity
at $z = 0.20258$ (see below for this choice). The profiles are complex with several components. 
From the \hi\ profiles (Ly$\beta$ and Ly$\delta$; Ly$\gamma$ is very 
noisy in STIS and is not used here, while in {\fuse}\ Ly$\gamma$ is contaminated, see Fig.~\ref{fig-lls}), 
we can decipher at least 5 blended components 
at about $-200$, $-160$, $-110$, $+50$, and $+205$ \km. In the $+50$ \km\ component, 
there is some evidence for at least another component near $0$ \km, but the absorption is 
too strong to be conclusive.

The combination of weak and strong metal lines allows us to
better explore the component structures. 
The \feii\ $\lambda$1144 and \feiii\ $\lambda$1122 profiles
show that the main absorption occurs at 0 \km. This component is also observed
in the profiles of \cii, \nii, \siii, \sii, and \oi. The latter is quite weak 
(see \S\ref{ssec-metal}), but is nevertheless detected at 3$\sigma$. As 
\oi\ is one of the best tracers of neutral hydrogen,
the positive detection of \oi\ $\lambda$1302 at $z = 0.20258$ (in 
agreement with the strong absorption in other 
species) sets the zero velocity point of the LLS. Examining the stronger lines (\siii, \siiii, \cii, \nii) shows other 
components at $-195$, $-162$, $-111$, $+46$, and $+95$ \km. Except for the latter, 
these components are also discerned in the \hi\ profiles, implying that the 
\hi- and metal-line components follow each other extremely well.

\subsection{\hi\ Column Densities}\label{ssec-hi}
The \hi\ Lyman series lines are detected from Ly$\alpha$ down to the Lyman Limit (see Figs.~\ref{fig-lls}
and \ref{fig-hi}). From the flux decrement observed at  the Lyman limit, we find
the optical depth $\tau_{LL}> 3$ and via $\tau_{LL}  = \sigma_{LL} N($\hi$)$ (where 
$\sigma_{LL} = 6.3 \times 10^{-18}$ cm$^{-2}$ is the photoionization cross-section for
hydrogen; Spitzer 1978 and Osterbrock 1989), we can place a firm lower limit on the \hi\ column density
(Fig.~\ref{fig-lls}): $\log N($\hi$) > 17.7$. In Fig.~\ref{fig-lls}, the thick solid shows the 
adopted flux levels blueward and redward of the break. As the continuum is 
not a straight line, we note that a change in the flux level redward of the break
of $\pm 0.1 \times 10^{-14}$ erg\,s$^{-1}$\,cm$^{-2}$\,\AA$^{-1}$  would change the limit on 
the column density by only about $\pm 0.1$ dex.  We estimate the flux blueward of the break
to be $<0.04\times 10^{-14}$ erg\,s$^{-1}$\,cm$^{-2}$\,\AA$^{-1}$. 

The total equivalent width of \lya\ is $1768 \pm 35$ m\AA, which would imply that 
$\log N($\hi$) \approx 18.8$ if it is on the square root part of the curve of growth. 
It is an upper limit because there is no evidence of damping wings in the Ly$\alpha$ profile (see also 
below). However, as we argue above, the Lyman limit system is confined to $>-80$ \km. Based on 
the weaker Lyman series lines (see Table~\ref{t-ew}), we integrate the \lya\ absorption from $-80$ to $+140$ \km
to find $W_\lambda \simeq 950$ m\AA, implying $\log N($\hi$) < 18.3$.

From these limits, the \hi\ column density is $17.7 < \log N($\hi$) < 18.3$. 
In order to refine the \hi\ column density, we also fitted the absorption Lyman series
lines with the Voigt component software of \citet{fitzpatrick97}. 
In the {\fuse}\ band, we assume a Gaussian instrumental spread function with a ${\rm FWHM_{inst}} = 20$ \km,
while in the STIS band, the STIS instrumental spread function was adopted \citep{proffitt02}. 
Note that our final fit does not use the Ly$\gamma$ line, but including this line would
not have changed our results because it is so noisy \citep[see Fig.~23 in][]{thom08}. 
For our adopted fit, we use the velocity-centroids of the components to 
those derived from the low ions as initial conditions (see \S\ref{ssec-kin} and column 2 in Table~\ref{t-hi})
but those are allowed to vary during the fitting procedure.  The Doppler parameter $b$ and column density $N$ 
in each component are also allowed to vary freely. The results are summarized in 
Table~\ref{t-hi} and the fits are shown in Fig.~\ref{fig-hi}. The reduced-$\chi^2$ 
for the resulting fit is 1.23. For the Lyman limit system at $z \equiv 0.20258$, we find 
$\log N($\hi$) = 18.15 \pm 0.25$, in agreement with the upper and lower limits derived
above. The velocities of the \hi\ derived from the fit are also in good agreement within $1\sigma$ 
from the velocities derived from the low ions (see columns 2 and 3 in Table~\ref{t-hi}).

\begin{deluxetable}{ccccc}
\tabcolsep=6pt
\tablecolumns{5}
\tablewidth{0pt} 
%\tabletypesize{\scriptsize}
\tablecaption{Kinematics and  Results from the \hi\ Profiles Fit \label{t-hi}} 
\tablehead{\colhead{$z$}   &\colhead{$v$}  &\colhead{$v_{\rm fit}$}   &  \colhead{$b$}  &  \colhead{$\log N$}  \\
\colhead{}   &\colhead{(\km)}   &\colhead{(\km)}   &  \colhead{(\km)}  &  \colhead{} \\
\colhead{(1)}   &\colhead{(2)}   &  \colhead{(3)}  &  \colhead{(4)} &  \colhead{(5)} 
}
\startdata
 0.201798	& $-195$      &  $  -187  \pm  11   $	&  $ 21 \pm  6   $  & $15.15 \pm    0.33 $  \\
 0.201930	& $-162$      &  $  -160  \pm  7    $	&  $ 12 \pm  8   $  & $15.20 \pm    0.31 $  \\
 0.202135	& $-111$      &  $  -113  \pm  1    $	&  $ 11 \pm  2   $  & $14.87 \pm    0.16 $  \\
 0.202580      &  $   0$      &  $    -1  \pm  10   $	&  $ 20 \pm  8   $  & $18.15 \pm    0.25 $  \\
 0.202765      &  $ +46$      &  $    26  \pm  11   $	&  $ 43 \pm 10   $  & $16.87 \pm    0.56 $  \\
 0.202961      &  $+95 $      &  $    85  \pm  13   $	&  $ 34 \pm  5   $  & $16.62 \pm    0.33 $  \\
 0.203398      &  $+204$      &  $   203  \pm  3    $	&  $  10 :       $  & $12.87 \pm    0.16 $  
\enddata
\tablecomments{Column (1): Redshifts determined from the low metal ions. \\
Column (2): Velocities of the components determined from the low metal ions; $z =  0.202580$
sets the zero-velocity. \\
Column (3): Velocities determined from the fit to \hi\ line profiles. \\
Column (4): Doppler parameter determined from the fit to \hi\ line profiles. A colon indicates that the 
value is uncertain (i.e. the error is of the order of the estimated value). \\
Column (5): Logarithmic of the column density (in cm$^{-2}$) determined from the fit to \hi\ line profiles.
}		
\end{deluxetable}

As the kinematics are complex and in particular the blending of the LLS with 
other \hi\ absorbers at higher absolute velocities could hide the presence of damping wings, 
we also tested the robusteness of the fit by forcing the value of $N($\hi$)$ at $z = 0.20258$ to 
be 18.5 dex and letting the $b$ in this component and $b,N$ in the other components to vary freely. 
For this value, the damping wings start to appear at positive velocities, 
confirming that the \hi\ column cannot be much greater than $10^{18.4}$ cm$^{-2}$. 

As the {\fuse}\ instrumental resolution  remains somewhat uncertain, we 
also investigated the results from the profile fitting using a Gaussian instrumental spread function with 
${\rm FWHM_{inst}} = 15$ and 25 \km, i.e. allowing for an error $\pm 5$ \km\ in FWHM$_{\rm inst}$. 
For all the components with $v\ge -120$ \km, an excellent agreement 
was found for $v,b,N$. Only in the components at $-195$ and $-162$ \km, we noticed a change in the column
density with the most negative component being the strongest when ${\rm FWHM_{inst}} = 25$  \km. However, 
the total column density in these two components was conserved. In view of the uncertainty, 
in the remaining of the text, we will  only consider the total column density, $\log N($\hi$) = 15.49 \pm 0.24$, 
of these components that we define as the absorber at $z \approx 0.2018$. 

Finally we also also used a curve-of-growth (COG) method to test our results for the LLS
and the absorber at $z \approx 0.2018$. As these \hi\ absorbers are strong, using different
analysis methods to determine $N($\hi$)$ is valuable since comparisons of the results 
provide insights about systematic error measurements.
The COG method used the minimization and $\chi^2$ error derivation approach
outlined by \citet{savage90}. The program solves for $\log N$ and $b$ 
independently in estimating the errors. In Table~\ref{t-ew}, we summarize our equivalent 
widths for these two absorbers. As the various components are strongly
blended together, for the LLS, we did not use Ly$\alpha$. For the absorbers at $z = 0.201798$ 
and $0.201930$, we combined these two absorbers (defined as the absorber at $z \approx 0.2018$).
The results of the COG are for the LLS, $\log N($\hi$) = 18.28\,^{+0.02}_{-0.33}$ ($b = 35 \pm 2$ \km), 
and for the absorber at $z \approx 0.2018$, $\log N($\hi$) = 15.59\pm 0.16$ ($b = 20 \pm 1$ \km). 
These values are quite consistent with those obtained from the profile fit. 
For our adopted \hi\ column densities, we take the mean between these two methods: 
$\log N($\hi$) = 18.22\,^{+0.19}_{-0.25}$ for the LLS and $\log N($\hi$) = 15.55\,^{+0.19}_{-0.24}$ 
for the absorber at $z \approx 0.2018$.

\begin{deluxetable}{ccc}
\tabcolsep=6pt
\tablecolumns{3}
\tablewidth{0pt} 
%\tabletypesize{\scriptsize}
\tablecaption{Equivalent Widths of the \hi\ transitions used in the COG \label{t-ew}} 
\tablehead{\colhead{$\lambda_{\rm rest}$}   &  \colhead{$\log(f\lambda)$} &  \colhead{$W_{\rm rest}$}  \\
\colhead{}   &  \colhead{(\km)} &  \colhead{(m\AA)}}
\startdata
  \cutinhead{$z = 0.20258$ -- $[-80,+140]^a$ \km}
 1025.7222  &   1.909   &	   $  740 \pm 37 $ \\
  949.7430  &   1.122   &	   $  605 \pm 69 $ \\
  937.8034  &   0.864   &	   $  570 \pm 61 $ \\
  930.7482  &   0.476   &	   $  550 \pm 56 $ \\
  926.2256  &   0.470   &	   $  536 \pm 56 $ \\
  923.1503  &   0.311   &	   $  515 \pm 59 $ \\
  920.9630  &   0.170   &	   $  531 \pm 62 $ \\
  919.3513  &   0.043   &	   $  506 \pm 61 $ \\
  918.1293  & $-0.072$  &	   $  467 \pm 70 $ \\
\hline
  \cutinhead{$z \approx 0.2018$ -- $[-235,-140]^a$ \km}
\hline
1215.6700  & 2.704 &  $ 381 \pm   19  $ \\
1025.7222  & 1.909 &  $ 266 \pm   19  $ \\
 949.7430  & 1.122 &  $ 137 \pm   39  $ \\
 937.8034  & 0.864 &  $ 140 \pm   30  $ \\
 930.7482  & 0.476 &  $ 54  \pm   32  $
\enddata
\tablecomments{$a:$ Velocity interval over which the equivalent widths were estimated.  
These intervals were defined using the uncontaminated weak \hi\ transitions. 
}		
\end{deluxetable}

Previous surveys of the low redshift IGM have considered this line of sight and this absorber, 
although none has considered the key information from the {\fuse}\ observations. 
\citet{thom08}  fitted the \hi\ column densities but only using
\lya, \lyb, and \lyg, resulting in a more uncertain fit
(note that there is a velocity shift between our results and 
theirs because they adopted for the zero velocity point from the centroids of the \ovi),
but in overall agreement with our results for the LLS ($\log N($\hi$) \sim 18.4$). \citet{danforth08} attempted to use the COG method
with \lya\ and \lyb\ equivalent widths and found $\log N($\hi$) = 15.14\,^{+0.21}_{-0.12}$ for the LLS, in contradiction
with our results and the lower limit derived from the strong break at the Lyman limit. 
Finally, \citet{tripp08} only estimated a lower limit on the column density of the LLS that is not really 
constraining ($\log N($\hi$) > 14.72$). For the absorber at $z \approx 0.2028$, $N($\hi$)$ values derived by
these groups are systematically smaller by about 0.5 dex than our adopted value although our results
overlap within about 1$\sigma$. The main difference is again that these groups did not use the 
weaker \hi\ lines available in \fuse. We note that the apparent optical depth method (see \S\ref{ssec-metal}) 
on the weaker $>3\sigma$ line (\hi\ $\lambda$937) yields $\log N_a = 15.43^{+0.13}_{-0.18}$,  in agreement within
1$\sigma$ with our adopted value.

\subsection{Metal Lines}\label{ssec-metal}

To estimate the column densities
and the Doppler parameters of the  atomic and ionic metal lines, 
we used the apparent optical depth \citep[AOD, see][]{savage91}.
In this method, the absorption profiles are converted 
into apparent optical depth (AOD) per unit velocity, 
$\tau_a(v) = \ln[I_{\rm c}/I_{\rm obs}(v)]$, where $I_{\rm obs}$,
$I_{\rm c}$ are the intensity with and without the absorption,
respectively.  The AOD, $\tau_a(v)$, is related to the apparent column
density per unit velocity, $N_a(v)$, through the relation
$N_a(v) = 3.768 \times 10^{14} \tau_a(v)/(f \lambda(\mbox{\AA})$)  ${\rm cm}^{-2}\,({\rm km\,s^{-1}})^{-1}$.
The total column density is obtained by integrating the profile, $N =
\int_{v_1}^{v_2} N_a(v) dv $.  For species that are not detected, we quote a 3$\sigma$
limit following the method described in \citet{lehner08}. 

Integration ranges for the equivalent widths and apparent column densities are listed in column 
2 of Tables~\ref{t-0202} and \ref{t-02026}. Table~\ref{t-0202} summarizes the total column
densities and equivalent widths of both absorbers at $z = 0.201798$ ($-195$ \km) and 0.201930
($-162$ \km). We use this approach because, except for \siiii,  the spectra are too 
noisy to reveal both absorbers. Table~\ref{t-02026} summarizes the results for the 
absorber directly associated to the LLS ($[v_1,v_2] = [-40,+30] $ \km)
or the absorbers associated to the LLS plus the absorption at higher positive velocities
($[v_1,v_2] \approx [-50,+130] $ \km). The reason for the latter interval is because 
\ovi\ and \nv\ cannot be separated into different components, and strong saturated lines (e.g. \siiii) 
do not show any structures in their profiles. As weaker transitions are only 
observed in the  $[-30,+30]$ \km\ interval, the column density in the LLS (0 \km) component 
dominates the total column density over  $[-50,+130]$ \km\ for the neutral, singly and doubly ionized 
species.

\begin{deluxetable}{lccc}
\tabcolsep=6pt
\tablecolumns{5}
\tablewidth{0pt} 
%\tabletypesize{\scriptsize}
\tablecaption{Measurements of the Metals at $z \approx 0.2018$ ($v \approx -190$ \km) \label{t-0202}} 
\tablehead{\colhead{Species}   &  \colhead{$[v_1,v_2]$} &  \colhead{$W_{\rm rest}$} &\colhead{$\log N$} \\
\colhead{}   &  \colhead{(\km)} &  \colhead{(m\AA)}&\colhead{}}
\startdata
%  \cutinhead{$z = 0.20257$	}
\cii\ $\lambda$1334	&	$[-220,-165]$ &	$46 \pm 23$	& $13.7 \,^{+0.1}_{-0.2}	$ \\
\cii\ $\lambda$1036	&	$[-220,-165]$ &	$41 \pm 9 $	& $13.68 \pm 0.10	$ \\
\ciii\ $\lambda$977	&	$[-220,-150]$ &	$148 \pm 42 $	& $> 13.45		$ \\
\nv\ $\lambda$1238	&	$[-210,-150]$ &	$29 \pm 6  $	& $13.21 \pm 0.08$ \\
\nv\ $\lambda$1242	&	$[-210,-150]$ &	$18 \pm 5$	& $13.27 \,^{+0.11}_{-0.15}$ \\
\ovi\ $\lambda$1031	&	$[-210,-150]$ &  $59 \pm 11$	& $13.81 \pm 0.09	$ \\
\ovi\ $\lambda$1037 	&	$[-210,-150]$ &	$27 \pm 11$	& $13.7  \,^{+0.1}_{-0.2}	$ \\
\siii\ $\lambda$1260 	&	$[-230,-170]$ & $58 \pm 12$	 & $12.61 \pm 0.10	$ \\
\siii\ $\lambda$1193 	&	$[-220,-160]$ & $25 \pm 10$	 & $12.6 \pm 0.2	$	 \\
\siiii\ $\lambda$1206 	&	$[-220,-145]$ &  $48 \pm 13$	 & $>13.00 $ \\
\siiv\ $\lambda$1393	&	$[-225,-145]$ &  $178\pm 50  $	 & $13.56\,^{+0.19}_{-0.34}  $ \\
\siiv\ $\lambda$1402	&	$[-225,-145]$ &  $105\pm 35  $	 & $13.54\,^{+0.13}_{-0.20}  $ \\
\aliii\ $\lambda$1854	&	$[-220,-150]$ &  $<189	$	 & $<13.05  $ 
\enddata
\tablecomments{Column density measurements were realized using the apparent optical depth by integrating
the profiles over the velocity interval $[v_1,v_2]$ (velocities are relative to $z=0.202580$).  
The column density of a feature detected in 
the 2--3$\sigma$ range is given only with one relevant digit.
``$<$" indicates a 3$\sigma$ upper  limit that was estimated over the velocity range observed in the absorption
of other species. ``$>$" indicates a lower limit. 
}		
\end{deluxetable}

\begin{deluxetable}{lccc}
\tabcolsep=6pt
\tablecolumns{5}
\tablewidth{0pt} 
%\tabletypesize{\scriptsize}
\tablecaption{Measurements of the Metals near $z \approx 0.2026$ \label{t-02026}} 
\tablehead{\colhead{Species}   &  \colhead{$[v_1,v_2]$} &  \colhead{$W_{\rm rest}$} &\colhead{$\log N$} \\
\colhead{}   &  \colhead{(\km)} &  \colhead{(m\AA)}&\colhead{}}
\startdata
\cii\ $\lambda$1036	&	$[-40,+30]$ &	$229 \pm 20$	& $> 14.31	$ \\
\cii*\ $\lambda$1037	&	$[-40,+30]$ &	$< 36	$	& $<13.51	$ \\
\nni\ $\lambda$1199	&	$[-40,+30]$ &	$< 39	$	& $<13.36	$ \\
\nii\ $\lambda$1083	&	$[-40,+30]$ &	$193 \pm 12$	& $>14.26	$ \\
\oi\ $\lambda$1302	&	$[-40,+30]$ &	$ 70 \pm 23$	& $14.17 \,^{+0.14}_{-0.22}$ \\
\siii\ $\lambda$1304	&	$[-40,+30]$ &	$182 \pm 46$	& $> 14.12	$ \\
\aliii\ $\lambda$1854	&	$[-40,+30]$ &	$< 31	$	& $<13.15	$ \\
\sii\ $\lambda$1259	&	$[-40,+30]^a$ &	$ 41 \pm 11$	& \nodata 	\\
\sii\ $\lambda$1253	&	$[-40,+30]$ &	$ 22 \pm 9$	& $14.1\,^{+0.2}_{-0.3}$ \\
\Siii\ $\lambda$1190	&	$[-40,+26]^b$ &	$ 84 \pm 12$	& $14.58 \pm\,^{0.10}_{0.06}$ \\
\feii\ $\lambda$1144	&	$[-40,+30]$ &	$ 51 \pm 14$	& $13.83 \pm 0.11$ \\
\feii\ $\lambda$1121	&	$[-40,+30]$ &	$ 25 \pm 11$	& $13.9 \,^{+0.2}_{-0.3}$ \\
\feii\ $\lambda$1096	&	$[-40,+30]$ &	$ 26 \pm 10$	& $14.0 \,^{+0.1}_{-0.2}$ \\
\feiii\ $\lambda$1122	&	$[-40,+30]$ &	$110 \pm 12$	& $14.46 \pm 0.07$ \\
\hline
\cii\ $\lambda$1334	&	$[-50,+130]$ &	$712 \pm 50$	& $> 14.60	$ \\
\cii\ $\lambda$1036	&	$[-50,+130]$ &	$444 \pm 20$	& $> 14.65	$ \\
\ciii\ $\lambda$977	&	$[-50,+130]^c$&	$592 \pm 77$	& $> 13.48		$ \\
\nii\ $\lambda$1083	&	$[-40,+130]$ &	$334 \pm 16$	& $>14.50	$ \\
\nv\ $\lambda$1238	&	$[-50,+130]$ &	$136 \pm 11$	& $13.85 \pm 0.05$ \\
\nv\ $\lambda$1242	&	$[-50,+130]$ &	$ 69 \pm 12$	& $13.88 \pm 0.08$ \\
\ovi\ $\lambda$1031 	&	$[-70,+150]$ &	$493 \pm 40$	& $14.93 \pm 0.10$ \\
\ovi\ $\lambda$1037 	&	$[-70,+150]$ &	$340 \pm 35$	& $14.95 \pm 0.05$ \\
\siii\ $\lambda$1304	&	$[-50,+130]$ &	$337 \pm 76$	& $> 14.45	$ \\
\siiii\ $\lambda$1206 	&	$[-62,+130]$ &	$646 \pm 18$	& $> 13.55	 $ \\
\siiv\ $\lambda$1393	&	$[-50,+130]$ &	$588 \pm 72$	& $> 14.12	 $ \\
\siiv\ $\lambda$1402	&	$[-50,+130]$ &	$501 \pm 67$	& $> 14.38	 $ \\
\svi\ $\lambda$933	&	$[-50,+130]$ &	$116 \pm 53$	& $13.7\,^{+0.2}_{-0.3}$ \\
\svi\ $\lambda$944	&	$[-50,+130]$ &	$ <188	$	& $< 14.04$ \\
\feiii\ $\lambda$1122	&	$[-50,+130]$ &	$144 \pm 18$	& $14.55 \pm 0.07$ 
\enddata
\tablecomments{Column density measurements were realized using the apparent optical depth by integrating
the profiles over the velocity interval $[v_1,v_2]$. The column density of a feature detected in 
the 2--3$\sigma$ range is given only with one relevant digit. When a feature is not detected 
at the 2$\sigma$ level (indicated by ``$<$"), we quote a 3$\sigma$ upper  limit that was estimated over 
the velocity range observed in the absorption of other species.  The ``$>$" sign indicates a lower limit. \\
$a:$  \sii\ $\lambda$1259 at 0 \km\ is  contaminated by \siii\ $\lambda$1260 at $-200$ \km\ relative to $z = 0.20258$.
$b:$  \Siii\ $\lambda$1190 is partially contaminated by \siii\ $\lambda$1190, which explains the smaller positive velocites. 
The larger positive error on $N$ takes this into account.  $c:$ \ciii\ may be partially contaminated as its absorption 
is observed well beyond $+130$ \km.
}		
\end{deluxetable}

{\em Absorber at  $z \approx 0.2026$} ($[-50,+130]$ \km):
Several absorption lines (\siiv, \siiii, \siii, \ciii, \cii, \nii) are saturated as evidenced 
by the core of the lines reaching zero-fluxes, and therefore 
the column densities are quoted as lower limits. We note that \citet{danforth08} estimated column  
densities for \siiv, \siiii, and \ciii, but we believe there is too little information to be able
to derive reliable column densities for these saturated absorption profiles. 
 The total apparent column densities of each \ovi\ doublet lines
are in agreement. Hence despite the absorption being extremely strong, these lines are essentially
resolved. The agreement also shows that the \ovi\ doublet lines are not contaminated
by unrelated lines.  The resulting weighted mean is $\log N($\ovi$) = 14.95 \pm 0.05$,
in agreement within 1$\sigma$ with previous estimates from \citet{tripp08,thom08}.
Results from \citet{danforth08} are $\sim$2$\sigma$ lower than other results, but
our equivalent widths are in agreement within 1$\sigma$. Saturation is also unlikely 
to play a role for \oi, \nv, \sii, \Siii, \svi, and \feii. \oi, \Siii, and \svi\ absorption is 
extremely weak. We can only integrate the  \Siii\ profile to $v = +26$  \km\ because it 
is blended with \siii\ $\lambda$1190. Using \siii\ $\lambda$$\lambda$1193, 1260, we note
that \siii\ $\lambda$1190 cannot contaminate \Siii\ at smaller velocities as no absorption
is observed in the stronger \siii\ lines at these velocities.  
The \oi\ $\lambda1302$ is a 3$\sigma$ detection and the relatively good alignment with other 
lines gives us some confidence that the line is not contaminated
by some weak \lya\ forest line. Unfortunately, \oi\ $\lambda$1039 is lost in a very a strong
absorption line. \sii\ $\lambda$1253 is barely a 2.4$\sigma$ detection and the stronger 
\sii\ line at 1259 \AA\ is contaminated with \siii\ $\lambda$1260 at $z \approx 0.2018$ ($v \approx -190$ \km)
(see below). For \feii, there are several transitions and within 1$\sigma$ error they 
are in agreement. We adopt the result from \feii\ $\lambda$1144 since only this
transition is detected above 3$\sigma$. 
The weak \nv\ doublet transitions are also in agreement within 
1$\sigma$. The resulting weighted mean for \nv\ is 
$\log N($\nv$) = 13.86 \pm 0.04$ \citep[within $\sim 1$--$2\sigma$ from previous estimates by][]{thom08,danforth08}. 
Finally, since $\tau_a(v=0) \simeq 1$ 
for the \feiii\ transition, this line might suffer from weak saturation. We assume for 
the remaining that saturation is negligible for this line but we keep this possibility in mind 
for our ionization models described below.  
We add that \feiii\ $\lambda$1122 is unlikely to be contaminated by an unrelated
intergalactic feature as this transition aligns very well with the other ones.

{\em Absorber at  $z \approx 0.2018$} ($[-220,-150]$ \km):
In contrast to the higher velocity absorbers, most of the absorption
profiles in this absorber are quite weak at the line center, $\tau_a < 1$. 
\cii\ $\lambda$1036 is a 4.6$\sigma$ detection whose column density
is in agreement with the 2$\sigma$ detection of \cii\ $\lambda$1334. 
The \nv\ doublet absorption is weak, the column densities
are consistent and both transitions are more than 3$\sigma$ detection, with 
a  resulting weighted mean $\log N($\nv$) = 13.23 \pm 0.07$ 
\citep[within $\sim 1\sigma$ from previous estimates by][]{thom08,danforth08}.
The \ovi\ $\lambda$1031 transition  is detected at 5.4$\sigma$, but \ovi\ $\lambda$1037
is only detected at 2.4$\sigma$. The column densities are, however, in agreement,
suggesting that \ovi\ $\lambda$1031 is not contaminated by intervening \lya\ forest line. 
Our estimate is consistent with those derived by \citet{danforth08} and \citet{thom08}. 
\ciii\ is saturated. \siiii\ has $\tau_a(v = -193) \sim 2.6 $, and we
therefore  quote only a lower limit in Table~\ref{t-0202}.
For \siii, only the transition at $\lambda$1260 is detected at the 3$\sigma$, 
the others are $>2\sigma$ detections. \siii\ $\lambda$1190 has a column density
0.3 dex higher than $\lambda$$\lambda$1193, 1260 and is likely contaminated. 
\sii\ $\lambda$1259 at $z=0.20258$ is directly blended with this feature (see above), 
but  both the \sii\ $\lambda$1253 line and the photoionization model  
presented in \S\ref{ssec-lls-phot} suggest that the strength of \sii\ $\lambda$1259
is too weak to be able to contaminate \siii\ $\lambda$1260.  The signal-to-noise
levels near the \siiv\ doublet are extremely low but nevertheless $\lambda$1393
is detected at 3.6$\sigma$ and $\lambda$1402 at 3$\sigma$. There is no evidence
of saturation in these lines and we adopt the weighted mean, $\log N($\siiv$) = 13.55 \pm 0.10$.
 
\section{Properties of the Absorbers}\label{sec-abs}
\subsection{Properties of the Absorber (LLS) at $z\approx 0.2026$}\label{sec-lls-prop}
\subsubsection{Metallicity}\label{ssec-lls-abund}
Using \oi\ and \hi, we can estimate the abundance of oxygen in the LLS without making any 
ionization correction (this is correct as long as the density is not too low or the ionization 
background not too hard, see below and Fig.~\ref{cloudy-lls}, and also Prochter et al. 2008). 
Using the column densities in \S\ref{ssec-hi} and Table~\ref{t-02026}, 
we derive $[$\oi/\hi$] = -0.7 \pm 0.3$. Throughout the 
text we use  the following notation ${\rm [X/H]} = \log N({\rm X}^i)/N({\rm H}^0) - \log({\rm 
X/H})_\odot$, where the solar abundances are from \citet{asplund06}. 
Because oxygen is generally a dominant fraction of the mass density in metals,
it is a valuable metallicity diagnostic.

Using the limit on \nni, we find $[$\nni/\hi$] < -0.7$ at 3$\sigma$, possibly suggesting
some deficiency of nitrogen relative to oxygen. The latter is consistent 
with either a nucleosynthesis evolution of N \citep[e.g.,][]{henry00} and/or a partial
deficiency of the neutral form of N because the gas is largely photoionized 
\citep{jenkins00,lehner03}.

\begin{figure*}[tbp]
\epsscale{1} 
\plottwo{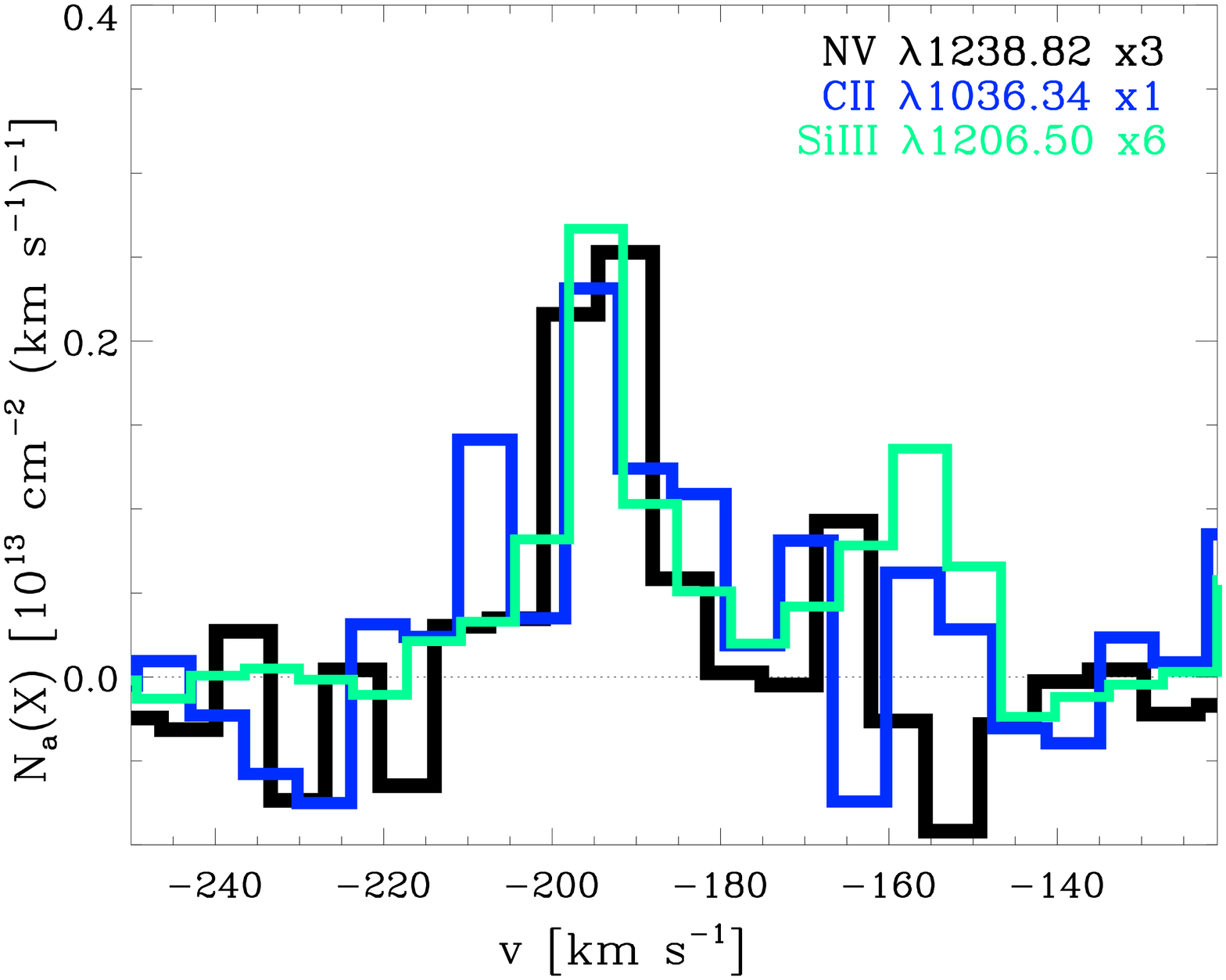}{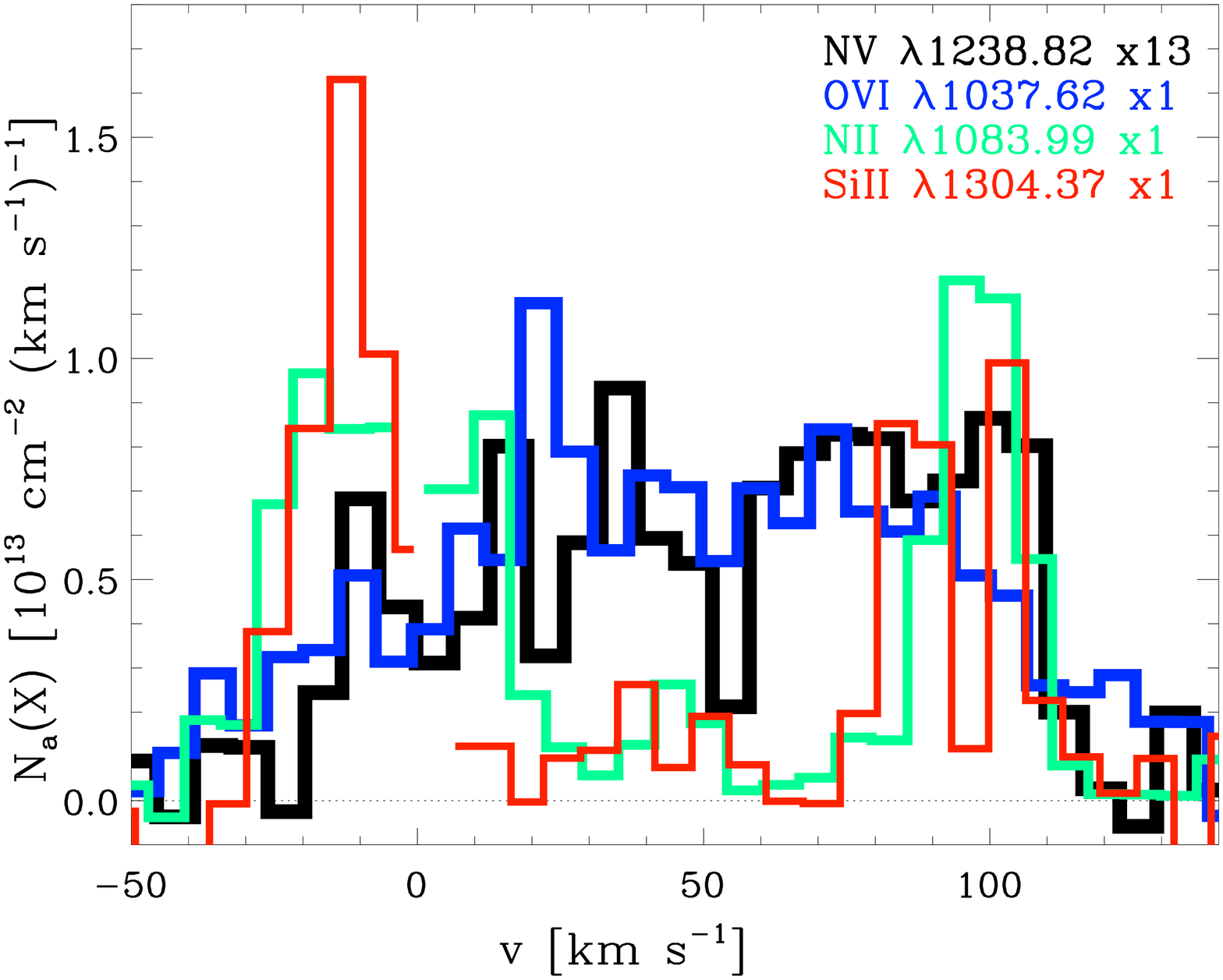}
\caption{Apparent column-density profiles of the metal ions relative to $z = 0.20258$. 
{\em Left panel}: AOD profiles for the absorber at $z \approx 0.2018$ ($v \approx -190$ \km). Note how well high, intermediate,
and low ions follow each other. 
{\em Right panel}: AOD profiles for the LLS.  Missing data points in 
the \nii\ and \siii\ profiles indicate that there is zero-flux at these velocities.  Note how well
\nv\ and \ovi\ follow each other with little structures in the profiles, while the low and intermediate
ions follow each other very well with clearly separated components at about $0$ and $+95$ \km. 
 \label{fig-aod}}
\end{figure*}

\subsubsection{Ionization}\label{ssec-lls-phot}
The LLS at $z\approx 0.2026$ shows absorption from atomic to weakly ionized species to 
highly ionized species. The singly (e.g., \cii, \siii) and doubly (e.g., \feiii) ionized 
species have very similar kinematics, suggesting that they probe the same gas 
(see Fig.~\ref{fig-norm}). In contrast, the \ovi\ and \nv\ profiles are broad, 
not following the same kinematics as the low ions. Fig.~\ref{fig-aod}
illustrates this further, where the $N_a(v)$ profiles of \siii, \nii, \ovi, 
and \nv\ are shown. Over the $[-50,+130]$ \km\ interval, \ovi\ and \nv\ 
$N_a(v)$  profiles follow each other very well but are quite different from 
the low-ion profiles. The stronger columns 
for the weakly ionized gas are between about $[-20,+20]$ \km, while 
at $\sim$45 and 100 \km\ the columns are much less important. 
For the high ions, the $N_a(v)$  profiles peak in $[+20,+100]$ \km\ interval.
This difference in the kinematics strongly hints at a multiphase gas, where
the high ions are produced by a different mechanism than the low ions
and/or the low and high ions are not co-spatial. 

In order to test these hypotheses, we first use a photoionization model to attempt to 
reproduce the observed column densities of the low ions and see if such a model 
can produce significant high-ion columns. We used the photoionization 
code Cloudy version C07.02 \citep{ferland98} 
with the standard assumptions, in particular that the plasma exists in 
an uniform  slab and  there has been enough time 
for thermal and ionization balance to prevail. 
We model the column densities of the different ions through a slab illuminated (on both 
sides) by the Haardt \& Madau (2005, in prep.) UV background  ionizing radiation field
from quasars {\em and}\ galaxies and the cosmic background radiation appropriate 
for the redshift $z = 0.2026$ ($J_\nu(912) = 5.4 \times 10^{-23}$ erg\,cm$^{-2}$\,s$^{-1}$\,Hz$^{-1}$\,sr$^{-1}$). 
We also assume a priori solar relative heavy element abundances 
from \citet{asplund06}.  We do not include the effects of dust on the relative abundances, 
although we consider this possibility a posteriori.  
We then vary the ionization parameter, $U = n_\gamma/n_{\rm H} =$\,H ionizing photon density/total
hydrogen number density [neutral\,+\,ionized], to search for models that 
are consistent with the constraints set by the column densities and $b$-values. 

\begin{figure}
\epsscale{1} 
\plotone{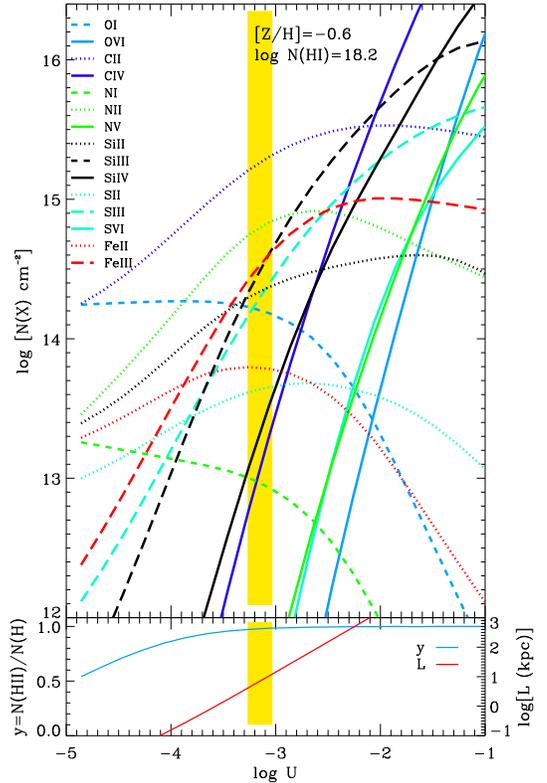}
\caption{{\em Top panel}: Predicted column densities for the Cloudy photoionization model of
the LLS  assuming a Haardt-Madau (galaxies+QSOs) spectrum at $z = 0.203$. The various lines show the models for 
each atoms or ions. Relative solar abundances are assumed. The yellow region shows a solution 
at $\log U =-3.15$ that fits the observations within about $\pm 0.1$ dex for \oi, \feii, \feiii. This model does, however, 
not produce enough highly ionized species.  Although \civ\ is not observed, we nevertheless show 
the model prediction as there might be future COS NUV observations of PKS\,0312-77.
{\em Bottom panel}: Variation of the neutral fraction and the path length ($L \equiv N({\rm H})/n_{\rm H}$).
For the solution that fits the observations, the fraction of ionized gas is about 98\% and the linear scale ranges
between 4 and 12 kpc. 
\label{cloudy-lls}}
\end{figure}

The results from the Cloudy simulations for the LLS are shown in Fig.~\ref{cloudy-lls} 
(see also the summary Table~\ref{t-02026m}). 
The Cloudy simulations stop running when $\log N($\hi$) = 18.2$ is reached. 
We adopted a metallicity $[{\rm Z/H}] = -0.6$ within the $1\sigma$ value 
derived using \oi/\hi. Smaller metallicities have difficulties in reproducing 
the column densities of \feii\ (and this would be exacerbated if Fe is depleted into
dust). In  Fig.~\ref{cloudy-lls}, the yellow region shows the solution 
$\log U = -3.15 \pm 0.10$ that reproduces the \Siii, \feiii, \oi, and \feii\ 
column densities within about 1$\sigma$ (see Table~\ref{t-02026m}). 
The limits of the other singly ionized species are consistent with this model. 
This range of $U$ values implies $\log N($\hii$)=   19.8 \pm 0.1 $, a density 
$\log n_{\rm H} = -2.5 \pm 0.1$, 
a temperature $T \simeq 1.1 \times 10^4$ K. The linear size of the absorber, $L \equiv N({\rm H})/n_{\rm H}$, 
is 4--12 kpc (a range of values not including the error on $N($\hi$)$ and $[{\rm Z/H}]$).

\begin{deluxetable}{lcccc}
\tabcolsep=6pt
\tablecolumns{5}
\tablewidth{0pt} 
%\tabletypesize{\scriptsize}
\tablecaption{Column densities and Abundances from the Cloudy photoionization model for the absorber at $z \approx 0.2026$ \label{t-02026m}} 
\tablehead{\colhead{Species}   &  \colhead{$\log N_{\rm obs}$$^a$}   &  \colhead{$[{\rm X^{+i}/H^0}]^a$} &  \colhead{$\log N_{\rm model}$$^b$}  &  \colhead{$[{\rm X^{+i}/H^0}]^b$}}
\startdata
\cii\	& $> 14.31     $	               &  $>-0.3$  		& $15.3,15.2 $ & $+0.7,+0.6 $ \\
\civ\	& \nodata		               &   \nodata		& $12.8,13.3 $ & $-1.8,-1.3 $ \\
\nni\	& $<13.36 $ 	    	               &  $< -0.7$  		& $13.1,13.1 $ & $-0.9,-0.9 $ \\
\nii\	& $>14.50 $ 	    	               &  $> + 0.5$  		& $14.8,14.9 $ & $+0.8,+0.9 $ \\
\nv\	& $13.86 \pm 0.04 $	               &  $-0.2 \pm 0.2$  	& $10.7,11.4 $ & $-3.3,-2.6 $ \\
\oi\	& $14.17 \,^{+0.14}_{-0.22}$           &  $-0.7 \pm 0.3$ 	& $14.3,14.3 $ & $+0.6,-0.6 $ \\
\ovi\	&  $14.95 \pm 0.05 $ 	               &  $+0.0 \pm 0.2$	& $7.2,9.8   $ & $-7.7,-5.1 $ \\
\siii\	&  $> 14.12	 $ 	               &  $>+0.4$	  	& $14.3,14.4 $ & $+0.5,+0.6 $ \\
\siiii\	&  $> 13.55	  $	               &  $>-0.2$		& $14.3,14.6 $ & $+0.5,+0.8 $ \\
\siiv\	&  $> 14.38	  $	               &  $>+ 0.6$	  	& $13.1,13.5 $ & $-0.7,-0.3 $ \\
\aliii\	&  $<13.15	  $	               &  $<+0.5$	  	& $12.6,12.9 $ & $-0.1,+0.2 $ \\
\sii\	& $14.1\,^{+0.2}_{-0.3}$               &  $+0.7 \pm 0.3$	& $13.7,13.7 $ & $+0.3,+0.3 $ \\
\Siii\	& $14.58 \pm\,^{0.10}_{0.06}$          &  $+1.2 \pm 0.2$	& $14.2,14.4 $ & $+0.8,+1.0 $ \\
\svi\	&  $13.7\,^{+0.2}_{-0.3}$              &  $+0.3 \pm 0.3$	& $10.5,11.2 $ & $-2.9,-2.2 $ \\
\feii\	& $13.83 \pm 0.11$	               &  $+0.1 \pm 0.2$ 	& $13.9,13.8 $ & $+0.2,+0.1 $ \\
\feiii\	& $14.46 \pm 0.07$	               &  $+0.8 \pm 0.2$ 	& $14.4,14.6 $ & $+0.7,+0.9 $ 
\enddata
\tablecomments{$a$: Estimated values for the absorber at $z \approx 0.2026$. 
$b$: Values from the Cloudy simulation presented in \S\ref{ssec-lls-phot} where the left and right hand-side
values are for $\log U = -3.25$ and $-3.05$, respectively. }		
\end{deluxetable}

Although the limit on \siiv\ is reproduced by this model, it is likely 
that there is some extra \siiv\ column not reproduced as 
the lines are so saturated for this doublet.
Both \nv\ and \ovi\ cannot be reproduced with this photoionization model by orders
of magnitude (see Table~\ref{t-02026m}). The column density of \svi\ is more uncertain, but
the photoionization also falls short to produce enough column for
this ion. Even if we consider only the velocity interval
$[-40,+30]$ \km, this discrepancy would still exist ($\log N($\ovi$)$ = 14.5
over $[-40,+30]$ \km). If non-thermal motions dominate the
broadening of the \nv\ and \ovi\ profiles, one could imagine that photoionization
with a large $U$ could reproduce the observed column densities if one invokes
that nitrogen is deficient (see Fig.~\ref{cloudy-lls}). 
This would require a much more diffuse gas ($n_{\rm H} < 6 \times 10^{-5}$ cm$^{-3}$) or an 
intense local source of hard ionizing radiation. The former model, in turn, is ruled out 
by the inferred size of the absorber, which implies a velocity sheer from Hubble broadening 
that far exceeds the observed velocity interval.

The broadenings of the \ovi\ and \nv\ profiles are large with 
$b($\ovi$) = 59 \pm 3$ \km\ and $b($\nv$) = 55 \pm 10$ \km, implying 
$T< (2$--$3) \times 10^6$ K. We note that \citet{tripp08} fitted the 
\ovi\ profiles with two components with $b = 48 \pm 7$ \km\ and $35 \pm 6$ \km, 
which is still consistent with $T<(2$--$3)\times 10^6$ K within the errors.  
\citet{danforth08} only fitted a single component to the \nv\ and \ovi\
profiles with $b = 63 \pm 6$ \km\ and $63 \pm 2$ \km, respectively. \citet{thom08} 
fitted the \ovi\ with a single component as well ($b = 68 \pm 3$ \km), but they fitted 
the \nv\ profiles with several components that follow somewhat the kinematics of \nii. 
In view of the excellent agreement between the \ovi\ and \nv\ $N_a(v)$ profiles
displayed in Fig.~\ref{fig-aod}, we are confident in our interpretation, i.e. the 
kinematics of \nv\ and \ovi\ must be similar (and different from the low ions). Our 
own inspection of the \ovi\ and \nv\ profiles shows that, since the data are quite noisy, 
a single- or two-component fit produce very similar reduced-$\chi^2$. It is quite possible
likely that there is more than one component in the high-ion profiles, but that does not rule out
the presence of $10^6$ K gas  
(if $b($\ovi$) = 32$ \km\ in the individual components and the broadening is mostly 
due to thermal motions, $T\sim 10^6$ K).

From the adopted column densities, the high-ion ratios are:  $N($\ovi$)/N($\nv$) \simeq 13$  and
$N($\ovi$)/N($\svi$) \simeq 11$--35. 
The high-ion ratios predicted by collisional ionization equilibrium 
(CIE) or non-equilibrium collisional ionization (NECI) models 
\citep{gnat07} are consistent with these values if $T\sim (3$--$10) \times 10^5$ K
as long as  $N($\ovi$)/N($\svi$) \ga 30$. Subsolar (down to $-0.6$ dex) to solar
$[{\rm N/O}]$ would be allowed in such models. A future estimate of the \civ\ column density
with the Cosmic Origins Spectrograph (COS) would further constrain this model. 
For $T<(2$--$3)\times 10^6$ K, the \hi\ column density
would be $<10^{14.2}$ cm$^{-2}$ in CIE or NECI (regardless of the metallicity), which may be 
accommodated for in the profile fitting presented in \S\ref{ssec-hi}. As both 
photoionized and collisionally ionized gas are present from $-50$ to $+130$ \km,
if the gas is co-spatial, it is multiphase. 

Although the \ovi\ and \nv\ profiles are broad, they do not appear that broad for such a high column 
density in the context of  models involving radiative cooling flows. \citet{heckman02} argue that
these models are able to naturally
reproduce the relation between $N($\ovi$)$ and $b($\ovi$)$ measured in various 
environments. However, using figure~1 in \citet{heckman02}, an \ovi\ absorber with $\log N ($\ovi$)=15$
should have $b \simeq 160$ \km\ in these models, a factor 3 larger than observed here. Furthermore 
the ratios $N($\ovi$)/N($\nv$) \simeq 25$--40  and $N($\ovi$)/N($\svi$) \simeq 250$--630 in these models 
are also quite different from the observed ratios 
\citep[for the \ovi/\svi\ ratio, see][]{lehner06}. Hence better physical models might involve 
shock-ionizations that heat the gas or interfaces between cool ($T\la 10^4$ K) and hot ($T >10^6$--$10^7$ K) plasmas
or a diffuse hot gas at $T \la 2\times 10^6$ K that has not yet had time to cool.    

We can gauge the fraction of the highly ionized gas relative to the neutral gas
by estimating  the amount of hydrogen in the highly ionized phase from 
$N_{\rm CI}($\hii$) = N($\ovi$)/(f_{\rm O\,VI} \, ({\rm O/H})_\odot 10^{[{\rm O/H}]})$ 
(the subscript ``CI" stands for ``collisional ionization"), where $f_{\rm 
O\,VI} = N($\ovi$)/N({\rm O})$ is the ionization fraction. For $T\sim (3$--$10) \times 10^5$ K, 
$f_{\rm O\,VI} =  0.2$--0.03 if the gas is in CIE or NECI. 
These $f_{\rm O\,VI}$ values imply  $\log N_{\rm CI}($\hii$) \simeq (19.0,19.8) - [{\rm O/H}] $. 
If the metallicity of the \ovi-bearing gas is similar 
to that of the LLS and $T \sim 3 \times 10^5$ K in the collisionally ionized gas, the \hii\ column density in the highly and photoionized 
gas would be about the same, implying $N_{\rm tot}($\hii$)\ga 10^{20}$ cm$^{-2}$ and
$N_{\rm tot}($\hii$)/N($\hi$) \sim 50$. If $T \sim 10^6$ K in the collisionally ionized gas, 
$N_{\rm CI}($\hii$)\ga 10^{20.5}$ cm$^{-2}$ and $N_{\rm tot}($\hii$)/N($\hi$) \sim 204$.
If the metallicity is much smaller than $-0.7$ dex, then $N_{\rm tot}($\hii$)/N($\hi$) \gg 54$--200, 
i.e. the amount of highly ionized gas could be indeed quite large.

\subsection{Properties of the Absorber at $z\approx 0.2018$}
As we discussed in \S\ref{sec-anal}, the absorbers at at $z = 0.201798$ and 
0.201930 are strongly blended with each other, and the \hi\ column densities in the 
individual components are somewhat uncertain. Therefore for our analysis
we treat these absorbers as a single one at  $z\approx 0.2018$ ($v \approx -190$ \km\ 
relative to the LLS), and adopt the total column density from our analysis in
\S\ref{ssec-hi}, $\log N($\hi$) = 15.55\,^{+0.19}_{-0.24}$.

The ionization properties of these absorbers appear quite different to the absorber
discussed above. The weak \nv\ absorption is narrow and well aligned with \siiii\ and 
\cii, and the AOD profiles for these ions follow each other extremely well (see Fig.~\ref{fig-aod}), 
suggesting that they arise in a single physical region. The $b$-value of \nv\ is $6.5 \pm 2.4 $ \km, 
implying $T< 7 \times 10^4$ K. The S/N levels near the \ovi\ absorption profiles are too low 
to derive any useful $b$-value. If the gas is collisionally ionized, it must be 
out of equilibrium and  has $Z>Z_\odot$ according to the calculations of  \citet{gnat07}. The ratio \nv/\ovi\,$\sim 0.3$  
can be produced in NECI models if $Z>Z_\odot$ and $T<3 \times 10^4$ K (assuming solar relative
abundances). This would require  a ratio of \siii/\siiv\,$\sim 5$--10 (consistent with the limit $>0.3$).
The ratio \cii/\siii\ would be a factor 2--5 too larger though. However, it is not clear how 
well these models tackle the low temperature regimes  where photoionization becomes important 
as well.

\begin{figure}
\epsscale{1} 
\plotone{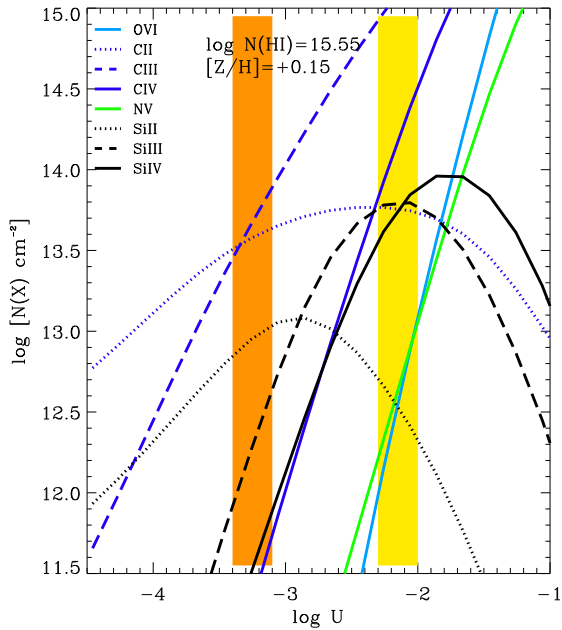}
\caption{Similar to Fig.~\ref{cloudy-lls} but for the absorber at  $z\approx 0.2018$ ($v\approx -190$ \km).
Note that here the metallicity is a free parameter and was varied to match the column 
densities of \cii, \siii, and \siiv. There are two possible solutions that are shown with the
orange and yellow regions. For the smaller $U$ solution, the intermediate and high ions need to  
be produced in non-equilibrium collisionally ionized gas. The other solution requires supersolar N and O relative to Si and C. 
For these solutions, the gas is about 100\% ionized and the linear scale is 
$\la 2$--5 kpc.
\label{cloudy-02018}}
\end{figure}

We therefore explore the possibility that the gas could be solely photoionized. 
Since the 3$\sigma$ upper limit on \oi\ $\lambda$1302 ($[{\rm O/H}] < + 2.0$)
is not useful, we let the metallicity be a free parameter in our Cloudy simulations. 
The calculations were stopped when the \hi\ column density reached $10^{15.55}$ cm$^{-2}$ and 
the metallicity was varied in order to reproduce the observed column densities of \cii\ and \siii\ 
(which are the most likely ions to be produced by photoionization in view of their low ionization
potentials).  Again, the metallicity must be high, $[{\rm Z/H}] = +0.15$ (and at least $[{\rm Z/H}] > -0.2$
when the error on $N($\hi$)$ are taken into account) in order to reproduce the column density of 
\siii\ and \cii\ (see Fig.~\ref{cloudy-02018}).  For $\log U \simeq -2.15$ (yellow region
in Fig.~\ref{cloudy-02018}), the observed column 
densities of \cii, \siii, and \siiv\ can be successfully reproduced within the errors.  
The lower limits on \ciii\ and \siiii\ also agree with this model. 
However, for this value of $U$, supersolar N and O abundances relative to C and Si
are required in order to fit \nv\ and \ovi. The orange region in Fig.~\ref{cloudy-02018}
shows another possible solution where $\log U \sim -3.15$: in this case the photoionization produces negligle column
densities for all the investigated species but \siii\ and \cii. A combination of photoionization 
and (non-equilibrium) collisional ionization may produce the observed ions for that solution.
 
Although it is not entirely clear which ionization processes dominate, for any models, 
supersolar abundances are required. The linear scale of the gas is also small ($<2$--5 kpc,
based on our photoionization model),
and any other (local) sources of ionizing radiation will only decrease
the size inferred for this absorber.
We note a Cloudy simulation where the UV background is dominated solely by QSOs would 
imply smaller $U$ but even a higher metallicity. The measurement of the \civ\ column density would
provide another constraint in order to test photoionization versus NECI models.

The origin of this absorber is therefore quite different from the LLS in view of the 
large metallicity difference, that despite the small redshift differences 
between these two absorbers. Since the gas is
100\% ionized, it is unlikely that the
sightline probes the gas from a galactic disk. It is more likely associated 
with some enriched material ejected from a galaxy. The size
of the absorber is also much smaller than usually derived in metal-line 
absorber with $\log N($\hi$)< 16$ \citep[e.g.,][]{prochaska04,lehner06}, 
further suggesting that it is closely connected to a galactic structure. 

\section{Physical Origin(s) of the Absorbers}\label{sec-gal}
\subsection{Summary from the Spectroscopic Analysis}
Before describing the galaxy survey in the field of view centered on 
PKS\,0312--77, we summarize the possible origins of the absorbers based on the
spectroscopic analysis. None of the absorbers originates
in the voids of the intergalactic medium where the influence from galaxies is negligible since
the gas in these absorbers is metal enriched. 
These absorbers therefore originate near galaxies, but unlikely through the disk of 
a spiral or an irregular galaxy since the gas is nearly 100\% ionized. 

The LLS (traced by the neutral and weakly ionized species) likely probes material
associated with the outskirts of a galaxy (e.g. cool galactic halo, or accreting, or outflowing material) 
or tidal material from an interaction between galaxies.
For another LLS at $z \simeq 0.081$  toward PHL\,1811 (also a largely ionized 
absorber with similar $N($\hi$)$), 
a possible origin was tidal debris \citep{jenkins03,jenkins05}. At $z = 0$, for the tidally disrupted
gas between the SMC and LMC, \citet{lehner08} show that this gas can be dominantly
ionized despite relatively high \hi\ column density as seen in the present absorber. 
Below we show that our galaxy survey also supports an interpretation involving  
tidal debris from a galaxy merger.  

There is also a strong absorption in the high ions near the LLS. We showed that
this gas cannot be photoionized but is collisionally ionized. The kinematics of the high 
ions also reveal little connection with the weak ions, except for the fact
that the velocity spread of the high- and low-ions (e.g., \cii, \siii, \siiii) 
are similar. In view of the difference in the kinematics and the ionization,  
it is likely that the \ovi\ arises in
a large volume of hot gas  in which the LLS is embedded. The CIE and NECI models
combined with the broadening of \nv\ and \ovi\ allow the gas to be at 
$T\sim (3$--$10) \times 10^5$ K.  Such high temperatures could reflect intragroup 
gas that is cooling, but we show in \S\ref{ssec-lls-phot} that the gas is unlikely in the 
process of cooling from a hotter phase. However, groups of late-type galaxies may 
produce cooler intragoup gas \citep{mulchaey96}, and we argue below that the \ovi\ absorber 
may be a tracer of such a plasma.  The strong \ovi\ 
absorber could also be the signature of a hot galactic corona around 
a galaxy or interacting galaxies. In this scenario, if the \ovi-bearing gas is 
in pressure equilibrium with the LLS, it implies
$n_{\rm O\,VI} = n_{\rm LLS} (T_{\rm LLS}/T_{\rm O\,VI}) = 0.4$--$1.2 \times 10^{-4}$ cm$^{-3}$,
where $n_{\rm LLS} \simeq 31 \times 10^{-4}$ cm$^{-3}$, $T_{\rm LLS}  = 10^4$ K, and 
$T_{\rm O\,VI} = (0.3$--$10) \times 10^5$ K (see \S\ref{ssec-lls-phot}). The cooling 
time is then $t_{\rm cool} \approx 0.6$--6 Gyr for the  \ovi-bearing gas. We showed in \S\ref{ssec-lls-phot} that
the properties of the \ovi\ absorber are not consistent with a radiatively cooling gas, so 
it seems more likely that the \ovi-bearing gas has been heated to
its peak temperature $3\times 10^5$ K, and hence  $t_{\rm cool} = 600$  Myr is more likely.
Below we argue that a galactic halo origin appears quite plausible.

Finally, the absorber at $z \approx 0.2018$ ($v \approx -190$ \km) with its supersolar metallicity could be
associated with a galactic wind or outflow from an enriched galaxy. For example, 
in our Galaxy, \citet{zech08} described a supersolar high-velocity cloud with
low $N($\hi$)$ ($=10^{16.50}$ cm$^{-2}$). The supersolar metallicity of this absorber implies a different 
origin than the LLS and suggests small-scale variation of the abundances if the absorbers are
co-spatial. 

\subsection{Las Campanas Observations}
\begin{figure*}
\epsscale{1.} 
\plotone{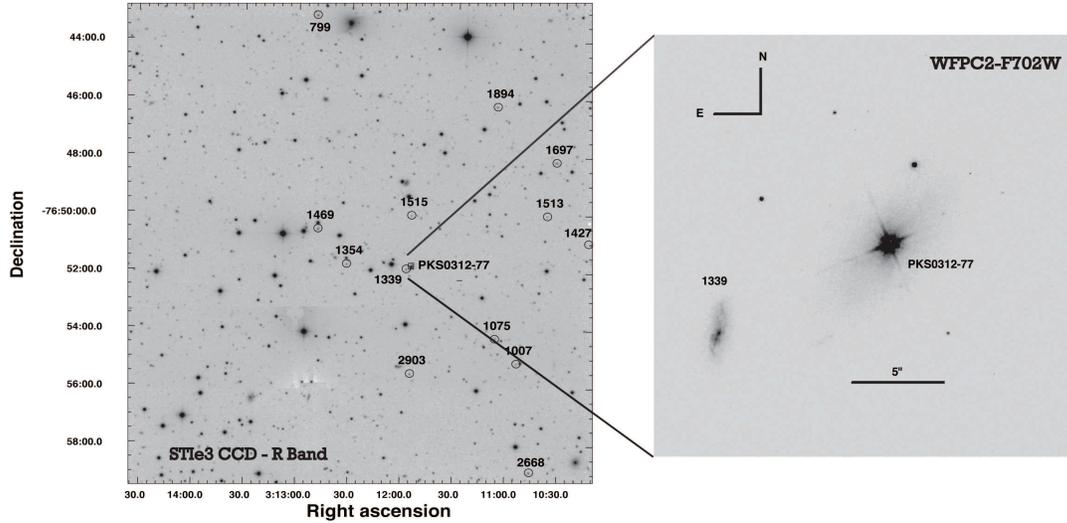} 
\caption{{\em Left:}  A $15.00 \arcmin \times 16.67 \arcmin$ cut from the Las Campanas galaxy survey. 
Positions of the galaxies and the QSO PKS\,0312--77 are shown. Only galaxies 
within $1200\,\mathrm{km\,s}^{-1}$ of the absorption system are marked (see Table~\ref{t-galsumm1}). 
{\em Right:} A zoom in on the region very near the QSO using a $36.4\arcsec \times 36.4 \arcsec$ cut of a {\em HST} 
WFPC2 image. The galaxy 1339 has only an impact parameter $ \rho = 38 h^{-1}_{70}$ kpc 
and velocity-offset of 16 \km\ from the Lyman limit absorbers. The morphology of this galaxy 
suggests a late-type interacting galaxy (see its optical spectrum in Fig.~\ref{fig-1339}). 
The outskirts of galaxy 1339 are the likely origin of the LLS and the absorber at $z \approx 0.2018$. 
\label{fig-galpos}}
\end{figure*}

To perform multi-slit spectroscopy, the field surrounding PKS\,0312--77 was first imaged in 
the R-band with Swope 1 meter telescope at Las Campanas Observatory. Six exposures of 3600 s 
were acquired on October 1 2002 and another 3 of 1800 s were acquired on  October 3 2002 with the SITe3 CCD in direct imaging mode
(pixel size 0.435\arcsec on a $2048 \times 3150$ array). These images were centered on 
PKS\,0312--77 and covered a field of view of $15\arcmin \times 32\arcmin$.
The conditions were photometric and the seeing fair. The exposures were taken
with a 10\arcsec\ dither pattern to account for bad pixels and to facilitate the construction
of a supersky flat. Full description of the data reduction can be found in \citet{prochaska06}. 
A $15.00 \arcmin \times 16.67 \arcmin$ cut is shown in Fig.~\ref{fig-galpos}.

In order to acquire the multi-object spectroscopy, follow-up observations were obtained with 
the Wide-Field CCD (WFCCD) spectrograph on the 2.5 meter Ir\'en\'ee du Pont telescope at Las
Campanas Observatory. We achieve $>90$\% completeness to $R \approx 19.5$ within
10\arcmin\ radius about  PKS\,0312--77. At $z\approx 0.2$, the survey covers a radius 
of $\sim 2$ Mpc and is 90\% complete for $L \ga 0.5 L_*$ galaxies.  A total 
of 132 spectra were taken in the field using 5 different slit masks, over the wavelength 
range 3600--7600 \AA. Two to three 1800 s exposures were taken per mask with a spectral 
resolution of 10 \AA\ and spectral dispersion of 2.8 \AA\ per pixel. The various steps
to reduce the data, separate galaxies from stars, and measure the galaxy redshifts
are fully described in \citet{prochaska06} and we refer the reader to this paper for 
more information. In total, we have confidently measured redshifts for 105 galaxies.  

\subsection{Results of the Galaxy Survey}\label{ssec-gal}
\subsubsection{A Galaxy and a Group of Galaxies}
In Table~\ref{t-galsumm1} we summarize the properties of the galaxies 
that are situated within  $\delta v = c (z-z_{\rm gal})/(1+z) = 1200$ \km\ 
from the absorbers (and $< 1100$ \km\ from the LLS, see Table~\ref{t-galsumm1}). Following \citet{prochaska06}, 
$\delta v$ is our first criteria for characterizing our sample of galaxies. 
This cutoff is somewhat arbitrary but allows for peculiar velocities 
in the largest gravitationally bound galaxies. We also do not impose 
a priori an impact parameter to allow for large scale structures. 
The last two columns in Table~\ref{t-galsumm1} 
list the quantities $L_C$ and $E_C$, which give information 
about the ``type" of the galaxy: early-type galaxies
have $E_C > 0.8$ and $L_C < 0.4$ and late-type galaxies have
$E_C < 0.8$ and $L_C > 0.4$. How these quantities are derived
is fully explained in \citet{prochaska06}. Finally, we adopted the absolute
magnitude of $L_*$, $M_R = -21.22$ at $z = 0.1$ \citep{blanton03}. When we report 
the $L_*$ value from other works in the literature, we corrected it
using the magnitude results from \citet{blanton03} and our adopted cosmology
if necessary.

\begin{deluxetable*}{rcccccccccc}
\tablewidth{0pc}
\tabcolsep=5pt
\tablecaption{Summary of Galaxies Neighboring Absorption Systems \label{t-galsumm1}}
\tablehead{\colhead{ID} & \colhead{$z_{gal}$}& \colhead{RA} & \colhead{DEC}  & \colhead{$R$} & 
\colhead{$L$} & \colhead{$\Delta \theta$} &\colhead{$\rho$}  &\colhead{$E_C$}  &\colhead{$L_C$}  &\colhead{$\delta v^a$}\\
\colhead{} & \colhead{}& \colhead{(J2000)} & \colhead{(J2000)}  & \colhead{} & 
\colhead{($L_*$)} & \colhead{($\arcsec$)} &\colhead{($h_{70}^{-1}$ kpc)}  &\colhead{}  &\colhead{} &\colhead{(\km)}}
\startdata
 1339 & 0.20264 &  03 11 57.90 & --76 51 55.68 & $ 19.2$ &  0.66  &  10.8 &    38 &  0.19 &  0.91 & $	16$ \\ 
 1515 & 0.20382 &  03 11 55.29 & --76 50 03.93 & $ 19.9$ &  0.36  & 106.8 &   356 &  0.03 &  0.68 & $  309$ \\ 
 1354 & 0.19874 &  03 12 31.87 & --76 51 46.31 & $ 19.3$ &  0.57  & 125.5 &   413 &  0.98 &--0.00 & $ -956$ \\ 
 1469 & 0.19822 &  03 12 48.42 & --76 50 33.63 & $ 18.5$ &  1.22  & 197.5 &   648 &  0.30 &  0.65 & $-1087$ \\ 
 2903 & 0.19821 &  03 11 54.98 & --76 55 33.36 & $ 19.1$ &  0.67  & 222.7 &   731 &  0.79 &  0.36 & $-1088$ \\ 
 1075 & 0.20288 &  03 11 06.85 & --76 54 18.78 & $ 18.3$ &  1.61  & 221.3 &   738 &  0.97 &--0.12 & $	76$ \\ 
 1513 & 0.20191 &  03 10 38.19 & --76 50 01.76 & $ 20.0$ &  0.32  & 283.9 &   942 &  0.09 &  0.68 & $-1025$ \\ 
 1007 & 0.19847 &  03 10 54.22 & --76 55 09.65 & $ 19.7$ &  0.41  & 287.5 &   943 &  0.45 &  0.61 & $ -168$ \\ 
 1427 & 0.20480 &  03 10 12.66 & --76 50 57.75 & $ 19.3$ &  0.62  & 353.2 &  1185 &  0.20 &  0.71 & $  553$ \\ 
 1697 & 0.20448 &  03 10 33.51 & --76 48 09.94 & $ 18.6$ &  1.14  & 355.1 &  1190 &  0.43 &  0.58 & $  473$ \\ 
 1894 & 0.20467 &  03 11 07.49 & --76 46 15.78 & $ 19.0$ &  0.80  & 372.2 &  1248 &  0.50 &  0.47 & $  521$ \\ 
 2668 & 0.20300 &  03 10 45.66 & --76 58 55.24 & $ 18.7$ &  1.09  & 486.1 &  1623 &  0.93 &  0.16 & $  103$ \\ 
  779 & 0.20398 &  03 12 49.91 & --76 43 09.44 & $ 18.5$ &  1.30  & 553.8 &  1855 &  0.95 &--0.20 & $  348$  
\enddata
\tablecomments{The galaxy summary is restricted to those galaxies
 within $1100\,\mathrm{km\,s}^{-1}$ of the absorption system.  The
impact parameter refers to physical separation, not comoving. Galaxy
redshifts were determined from fitting the four SDSS star and galaxy
eigenfunctions to the spectra \citep[see][]{prochaska06}. The coefficient
of the first eigenfunction $E_C$ and a composite of the last three
eigenfunctions $L_C$ are used to define galaxy type. Early-type
galaxies have $E_C > 0.8$ and $L_C < 0.4$, while late-type galaxies
have $E_C < 0.8$ and $L_C > 0.4$. $a$: Velocity separation between the galaxy redshifts and the LLS at $z=0.20258$.} 
\end{deluxetable*}

The left-hand side of Fig.~\ref{fig-galpos} shows the galaxy and the QSO 
positions for the galaxies $\delta v \le 1100 $ \km\ from the LLS. The galaxy 1339
stands out in view of its proximity to the QSO. For this galaxy, 
 $|\delta v| <100 $ \km\ relative to the LLS and \ovi\ absorption system
and the impact parameter is $38 h^{-1}_{70}$ kpc, making this galaxy the most
likely host of the LLS and \ovi\  absorber. The next 
closest galaxy (\#1515) is already at  $358 h^{-1}_{70}$ kpc and $\delta v > 200 $ \km.
On the right-hand side of Fig.~\ref{fig-galpos}, we zoom in on the region 
very near the QSO using a $36.4\arcsec \times 36.4 \arcsec$ cut of a {\em HST} 
WFPC2 image using the F702W filter (the observations were obtained by PI M. Disney (program 6303), 
and consist of  four exposures totaling 1800 s; we used standard procedures to reduce the WFPC2 
images). The {\em HST}\ image goes deeper than our galaxy redshift survey, but except
for galaxy 1339 ($0.7 L_*$), the only galaxies within 100 kpc are less than $0.1 L_*$ 
(assuming they are at redshift $z \sim 0.203$). Dwarf galaxies are known to
produce outflows \citep[e.g.][]{martin99} and \citet{stocke06} also argue that it
is quite likely that many of the responsible galaxies via their outflows and halos 
for the \ovi\ absorbers may be $<0.1 L_*$ galaxies. However, since the present \ovi\ absorber is the
strongest yet discovered and the linear-scales of the absorbers are quite small ($<12$ kpc, 
see \S\ref{sec-abs}), a scenario where $<0.1 L_*$ galaxies would be responsible for
these strong absorbers does not appear compelling (see also below). 
Future deeper searches below $0.1 L_*$ will be needed to uncover the true impact 
of  $<0.1 L_*$ galaxies on \ovi\ absorbers. 

Beyond 300 kpc,  there are 5 more galaxies that have  $|\delta v| <300 $ \km\
relative to the absorbers at $z \approx 0.2026$ to 0.2030, and 
13 with  $|\delta v| \la 1100 $ \km. These galaxies have large impact parameters of 
$358 \le \rho \le 1856 h^{-1}_{70}$ kpc, consistent with a group of galaxies. 
Intragroup gas from this group could also be responsible for the \ovi\ 
absorber. Below, we first review the properties of galaxy 1339 and then address 
the possible origins of the absorbers.

\subsubsection{Properties of Galaxy 1339}
\begin{figure}
\epsscale{1.} 
\plotone{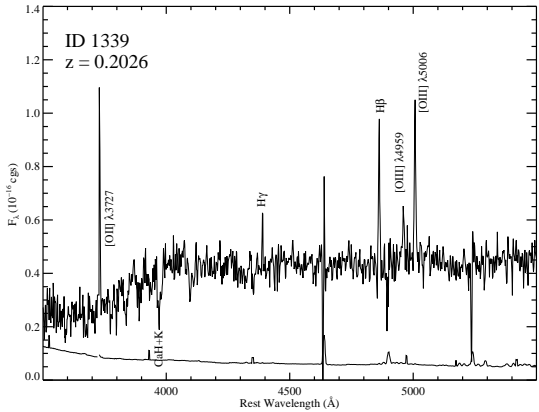}
\caption{Spectrum of galaxy 1339, the likely host of the Lyman limit system
at $z = 0.20258$. 
\label{fig-1339}}
\end{figure}

The appearance of galaxy 1339 is consistent with a late-type galaxy derived from 
the $E_C$ and $L_C$ parameters. A close inspection to the galaxy suggests
that it has been subject to a collision or an interaction
with another galaxy as there appear to be two bulges separated by about 0.4\arcsec\ 
(or $\sim 1 h^{-1}_{70}$ kpc), indicating some disruption in this galaxy.  
In Fig.~\ref{fig-1339}, we show the spectrum of galaxy 1339.  
The emissions of [\oii], [\oiii], and H$\beta$ with the property
$W([$\oii$_{\lambda3727}])/ W([$\oiii$_{\lambda5907}]) \approx 1.3$  and 
$W([$\oii$_{\lambda3727}])/   W({\rm H}\beta) \approx 2.1 $ 
closely  mimic the properties of a  Sc pec galaxy \citep[e.g., NGC\,3690,][]{kennicutt92a,kennicutt92b} 
or Sm/Im pec galaxy \citep[e.g., NGC\,4194,][]{kennicutt92b} that 
is undergoing a close galaxy interaction or merger, confirming our 
visual inspection of Fig.~\ref{fig-galpos}.

Galaxy 1339 is very unlikely to have interacted with another galaxy in a manner like 
the Galaxy is currently interacting with the LMC and SMC. If that was the case, 
one would expect to observe a $>0.1 L_*$ galaxy within a few tens of kpc of galaxy 1339. 
As the {\em HST}\ images does not reveal any other potential $\ga 0.1 L_*$
galaxy, the only possibility would be that a galaxy is hidden by the glare of
the QSO. We have explored this possibility by subtracting the QSO using 
a mask to exclude the saturated pixels in the center and surrounding objects.
The residual image reveals no serendipitous galaxy or fine structure.  Therefore, galaxy 1339
is likely the result of a galaxy merger; perhaps the close interaction between the
SMC and LMC will lead these galaxies to a similar fate before being cannibalized by
the Milky Way. 

The metallicity of the galaxy is an important ingredient to know for comparing the abundances
in the absorbers and the likely host galaxy.  
We can estimate the gas-phase oxygen abundance for
galaxy 1339 using the ratios of strong nebular emission lines \oii\
$\lambda$3727, H$\beta$, and \oiii\ $\lambda\lambda$4959, 5007
\citep[e.g.,][]{pagel78}.    We assume that the \hii\ regions covered
by the spectroscopic slit are chemically
homogeneous, and we adopt the calibration between oxygen abundance, 
$12 +\log({\rm O/H})$, and the strong line ratios $R_{23}$ and $O_{32}$ based on the
photoionization models of \citet{mcgaugh91} as described in \citet{kobulnicky99}.  
We use both the method of flux ratios (not
corrected for reddening) and the method of emission line equivalent
widths ratios introduced by \citet{kobulnicky03} that is more
robust against reddening;  both yield very similar results.  We find
$12 +\log({\rm O/H}) = 8.90 \pm 0.05$.  By comparison, the line ratios of the
Orion nebula \citep{baldwin91}, which might be taken as
representative (for emission line studies) of the solar neighborhood, yield $12 +\log({\rm O/H}) \simeq 8.75$ with a
dispersion of about 0.04 dex among multiple sightlines.  Hence, the emission lines from galaxy 1339
arise in a region approximately 0.15 dex more metal rich than the solar vicinity in the Milky Way.   

Using the H$\beta$ line, we can also roughly estimate the star-formation rate (SFR) in galaxy 1339. From our 
optical spectrum, we measure $F_{\lambda}({\rm H}\beta) \simeq 3.3 \times 10^{-16}$ erg\,s$^{-1}$. 
At $z=0.203$, this implies $L({\rm H}\beta)\simeq 4.1 \times 10^{40}$ erg\,s$^{-1}$ for our adopted cosmology.  
Assuming no extinction and a standard luminosity ratio $L({\rm H}\alpha)/L({\rm H}\beta)=2.86$, we find
$L({\rm H}\alpha) \approx 4.1 \times 10^{40}$ erg\,s$^{-1}$. Using \citet{kennicutt89} and assuming an extinction 
of about 1 magnitude, the SFR is $2\pm 1 $ $M_\odot$\,yr$^{-1}$, which is roughly consistent with 
a $\sim$$L_*$ galaxy. Hence while galaxy 1339 is not a starburst, it nevertheless sustains star formation, allowing
the possibility of stellar feedback, which can produce galactic outflows.  We also note that
it is quite possible that burst of star formation may have occurred several tens to hundreds of Myr before the
observed epoch, leaving open the possibility of violent star formation and mass ejection in the past. 
This conjecture is supported if galaxy 1339 was subject to a galaxy merger as these events 
are known to create new burst of star-formations \citep{larson78,barton07,demello07}. 

\subsection{Origins of the LLS and $z\approx 0.2018$ Absorber}
In view of its properties and impact parameter, galaxy 1339 is a very likely 
candidate for the origin of the LLS and the absorber at $z\approx 0.2018$ ($v \approx -190$ \km). The supersolar metallicity of that galaxy
is (remarkably) the same as the one derived for the absorber at $z\approx 0.2018$. 
The high velocity of the absorber relative to the galaxy velocity  ($\delta v = 178$--211 \km) fits 
nicely in a scenario involving a galactic wind or outflow from the galaxy. As we show above that 
star formation is occurring in that galaxy, galactic feedback involving supernovae and stellar winds is 
quite possible.  According to the recent simulations of 
feedback within cosmological  models, a galactic wind may travel on physical distances of 60--100 kpc
\citep{oppenheimer08a}, consistent with the impact parameter of 38 kpc for galaxy 1339. In these models, even 
$\sim$$L_*$ galaxies may produce such outflows. 

The velocity offset between the LLS and the galaxy is small ($|\delta v | \simeq 16$ \km), suggesting 
it is bound to the galaxy if projection effects are negligible. Because there is strong 
evidence that galaxy 1339 is the result of a galaxy merger, it appears reasonable to hypothesize 
that the LLS traces some leftover debris from the merger.  It is interesting to note
that the physical distance of the LLS from galaxy 1339, linear size, and \hi\ column
density are in fact very similar to recently found \hi\ clouds in the halo of M\,31 \citep{thilker04}
and in the M81/M82 group \citep{chynoweth08}. These clouds are thought to be the extragalactic
counterparts of the high-velocity clouds (HVCs) observed in the Galactic halo \citep[e.g., see review by][]{wakker01}. 
Tidal disruption is also the most obvious origin considered for the \hi-halo clouds near 
M31 and M81/M82 galaxies \citep{thilker04,chynoweth08}, especially for the M81/M82 group, which
in contrast of M31 is undergoing a strong galactic interaction. Yet neither for the present absorber
nor for these other nearby galaxies, other scenarios can be entirely rejected (e.g., 
condensation of galactic halo material or even outflows several tens or hundreds of Myr before
the observed epoch).
 
The metallicity of the LLS is quite different from that of the absorber at $z\approx 0.2018$ 
and the galaxy itself. First, we note that the metallicity of galaxy 1339 is unlikely to 
be homogeneous, especially if it resulted from a recent merger. The different metallicity may 
arise if the galaxies that merged had different metallicity and/or the leftover debris were 
mixed   with more pristine gas.  The difference in metallicity is also indicative of 
poor metal mixing on galactic-scale structure of tens of kpc. Evidence for poor metal mixing 
in the $\la 100$ kpc halo of galaxies is also observed at lower redshift in our own Galactic halo  
\citep[e.g.,][]{wakker01,collins03,tripp03}, in the Magellanic system \citep[][N. Lehner et al. 2009, in prep.]{gibson00,lehner08}, and at higher $z$ 
in other LLS \citep{prochter08}, suggesting it is a systematic property of galactic halos \citep[see also][]{schaye07}. 

We noted above that the \ovi-bearing and the LLS could be in pressure equilibrium
if they are spatially coincident (see \S\ref{sec-ovi-orig}), then the LLS could be pressure
confined by coronal gas traced by the high-ion absorber. Even if the LLS is not 
pressure confined, the lifetime of such a cloud with the properties derived in 
\S\ref{ssec-lls-phot} would  be about 0.3--0.8 Gyr using the expansion-time equation 
given by \citet{schaye07}. It is reasonable, therefore, to associate the LLS with
debris from a merger that is just ending and has been happening on the timescale of $\la 1$ Gyr.

\subsection{Origin(s) of the Strong \ovi\ Absorber}\label{sec-ovi-orig}
In the picture presented above, the \ovi/\nv-bearing gas may represent a hot, collisionally 
ionized gas about galaxy 1339 as the \ovi\ absorption revealed in Galactic halo sightlines
\citep{savage03,sembach03}  or about interacting galaxies like the \ovi\ absorption observed toward
the LMC or SMC \citep{howk02,hoopes02,lehner07}.  However, there is also a 
group of galaxies with $358 \le \rho \le 1856 h^{-1}_{70}$ kpc. It is therefore possible that the 
\ovi\ absorption is so strong and broad because it probes the large-scale gravitational 
structures inhabited by the galaxies summarized in Table~\ref{t-galsumm1}. 
We explore now if the properties of the \ovi\ absorber compared to other absorbers and to 
results from cosmological simulations allow us to differentiate between these two scenarios. 

Many aspects of feedback are still too complex to reliably model (e.g., metal cooling, non-equilibrium 
ionization effects) and simulations often lack the spatial and mass resolutions to resolve the
supernova environment. Nevertheless, recent simulations attempt to model galactic outflows 
in a cosmological context, showing that feedback is a necessary ingredient and galactic 
winds are required to match the low-column \ovi\ absorber density \citep[e.g.,][]{cen06}. 
\citet{oppenheimer08} specifically investigated the origin(s) of the \ovi\ absorbers in 
{\sc Gadget-2} cosmological simulations that use a variety of physics inputs and include
galactic outflows with various strengths.
They found that strong \ovi\ absorbers are usually collisionally 
ionized, but also are found in multiphase gas and may be misaligned relative to the
low ions as the present \ovi\ absorber. In their models, strong \ovi\ absorbers trace
the outskirts of the halos, while absorbers with $\log N($\ovi$) \ga 15$ trace metals
in galactic halo fountains within the virial radius. 
In their models, galactic winds travel distances of $\la 100$ kpc, but 
do not escape the galaxy halo and fall back down in a ``halo fountain''  on recycling timescales 
$\le 2$ Gyr \citep{oppenheimer08a}. 
According to figure~12 in \citet{oppenheimer08}, $\ga 0.3 M_*$ galaxies may recycle absorbers with
$\log N($\ovi$) \sim 15$ within $\sim$1 Gyr. Above we argue that the absorber 
$z \approx 0.2018$ might be evidence for a galactic wind, supporting further this interpretation. 
Therefore, the present strong \ovi\ absorber supports findings of  \citet{oppenheimer08}. 

From an observational perspective, the properties of the \ovi\ absorber
are quite similar to another very strong \ovi\ absorber for which galaxy information 
exists: Toward PKS\,0405--123, a strong \ovi\ absorber with $\log N($\ovi$) = 14.78$ 
and $\log N($\nv$) = 13.89$ was detected at $z= 0.16710$ 
\citep{chen00,prochaska04,williger06}. The full velocity extents of the main \ovi\ 
and \nv\ absorption are also quite similar with $\Delta v  \approx 150$ \km. 
Both absorbers exhibit two distinct phases: (i) a photoionized gas at $T\sim 10^4 $ K and (ii)
a hot ($T\sim 2\times 10^5$--$10^6$ K), collisionally ionized gas associated with \ovi, \nv, and \svi\ absorption. 
For both absorbers, velocity offsets from velocity of the host galaxy 
are small, suggesting that both gas phases are bound to the galaxy. 
The properties of the galaxies have, however, some key differences. First,
in the field about PKS\,0405--123, the likely host galaxy is much brighter, with  $ L \sim 3.4 L_*$.
Secondly, there is no evidence for a group of galaxies within 3 Mpc of PKS\,0405--123. 
In view of these properties,  a galactic halo origin rather than intragroup medium was strongly favored for 
the \ovi\ absorber at $z \simeq 0.1671$ toward  PKS\,0405--123
\citep{prochaska06}.

Other strong \ovi\ absorbers have been found \citep[e.g.,][]{tripp08},  but they are rare,
and even more so with close spectroscopically identified galaxies.
In their  survey of the very local Universe ($z<0.017$), \citet{wakker08} reported a
strong \ovi\ absorber with $\log N($\ovi$)= 14.63 \pm 0.15$. This
absorber has a $2.1 L_*$ galaxy  at $\rho \simeq 62 h^{-1}_{70}$ kpc
and with $\delta v = 100$ \km. This suggests that strong \ovi\ absorbers are generally 
found near $>L_*$ galaxies. On the other hand, not 
all galaxies with $\rho \la 100  h^{-1}_{70}$ kpc have strong \ovi\ absorption: 
the LLS at $z = 0.08$ has two S0 galaxies with $\rho \simeq 34$ and $ 87 h^{-1}_{70}$ kpc
but no \ovi\ absorption \citep{jenkins05}; at $z<0.02$, several \hi\ absorbers have no \ovi\ or weak 
\ovi\ absorption for galaxies with  $\rho \la 100  h^{-1}_{70}$ kpc \citep{wakker08}, 
and at $z < 0.15$, \citet{stocke06} found two-thirds of the \ovi\ non-detections are found 
within $1 h^{-1}_{70}$ Mpc of the nearest galaxy.  The absence 
of strong \ovi\ absorption within 100 kpc of a galaxy is, however, not
entirely surprising: the detection rate of \ovi\ absorption 
(irrespective of its strength) at such an impact parameter is less than 50\% \citep{stocke06,wakker08}.
The detection probability can be understood if the \ovi-bearing gas is patchy
and distributed in complicated sheet-like structures \citep{howk02a},
and because the finite distances that metals can reach once ejected from 
galaxies  \citep{tumlinson05,stocke06}.

\begin{deluxetable*}{lcccccccc}
\tablewidth{0pc}
\tabcolsep=5pt
\tabletypesize{\scriptsize}
\tablecaption{Variety of LLS and ``Associated'' \ovi\ Absorbers  at  $z \la 0.2$\label{t-lls}}
\tablehead{\colhead{Sightline} & \colhead{$z_{\rm LLS}$} & \colhead{$\log N($\hi$)$}& \colhead{$\log N_{\rm OVI}$}& \colhead{$[{\rm Z/H}]_{\rm LLS}$}& \colhead{Likely host galaxy}& \colhead{Possible Origin(s)}& \colhead{Possible Origin(s)} & \colhead{Ref.}\\
\colhead{} & \colhead{}&  \colhead{}&\colhead{}& \colhead{}& \colhead{$\delta v, \rho, L$}& \colhead{LLS}& \colhead{\ovi}& \colhead{} \\
\colhead{} & \colhead{}&  \colhead{}&\colhead{}& \colhead{}& \colhead{(\km, kpc, $L_*$)}& \colhead{}& \colhead{}& \colhead{} 
 }
\startdata
PKS\,0405--123    & 0.16710& $16.5 $& $ 14.8$  & $-0.3$       & $-15,108,3.4$	&  galactic halo	& galactic halo			& 1        \\ 
PKS\,1302--102    & 0.09847& $17.0 $& $ 14.0$  & $\la -1.6$   & $-354,65,0.2$	& outflow/inflow?	&  outflow/inflow?		& 2\\ 
PHL\,1811         & 0.08092& $18.0 $& $ <13.2$ & $-0.2$       & $+36,34,0.5$   	&  tidal debris/wind	& \nodata			& 3\\ 
PKS\,0312--77     & 0.20258& $18.3 $& $ 15.0$  & $-0.6$       & $+16,38,0.7$	& merger debris		& galactic halo/intragroup	& 4    
\enddata
\tablecomments{
PKS\,0405--123: There are two galaxies (the other is $\la 0.1 L_*$) within 108 kpc. All the other galaxies are at $>3$ Mpc (sensitivity of
the survey  is $0.1 L_*$). \\
PKS\,1302--102: The large $\delta v$ requires extremely large outflow/inflow for a small galaxy. Ten ($0.1L_*\la L\la 5L_*$) 
galaxies are found  at $\rho  \la 800$ kpc and with $100 \la |\delta v| \la 600$ \km, which might suggest the intragroup medium 
as the origin of the LLS and \ovi\ (sensitivity of the survey is $\sim 0.2 L_*$). \\ 
PHL\,1811: Another $0.5 L_*$ galaxy is found with $\delta v = 146$ \km\ and at  87 kpc. \\
PKS\,0312--77: See this paper. \\ 
{\em References:} (1) \citet{prochaska04,prochaska06}; (2) \citet{cooksey08}; (3) \citet{jenkins03,jenkins05}; (4) this paper. 
  }
\end{deluxetable*}

While \citet{oppenheimer08} relate collisionally ionized \ovi\ absorbers 
to HVCs or IVCs observed in the Milky Way halo, such strong \ovi\ 
is not observed in the {\em FUSE}\ \ovi\ survey of the Galactic halo \citep{wakker03,savage03,sembach03}. 
Strong \ovi\ absorption is generally related to the ``thick" disk of the Galaxy, 
but thick disk material cannot have been probed  by this sightline. 
On the other hand,  the \ovi\ might be so strong because it probes the
halo of a galaxy merger, which possibly produced in the past a strong burst 
of star formation, and hence strong galactic feedback. As we have alluded to above, 
galaxy 1339 may be an evolved system of the fate awaiting  the SMC-LMC system, 
two sub-$L_*$ galaxies. For the \ovi\  absorption toward the LMC stars, \citet{howk02} and \citet{lehner07} 
argue that the \ovi\  absorption probes a hot halo and feedback 
phenomena associated with the LMC. The LMC \ovi\ column densities are
generally smaller than 14.5 dex \citep[see summary table~7 in][]{lehner07}. However, 
these sightlines only pierce one side of the halo of the LMC as the background
targets are stars. In the SMC, one line of sight has $\log N($\ovi$) \approx 14.9$
and several have $\log N($\ovi$) > 14.6$ \citep{hoopes02}. The enhancement is
related to the stellar activity within the SMC and it is not clear how much of the 
\ovi\ absorption arises in the SMC halo versus the SMC disk. Yet, it is not outside 
of the realm of possibility that a sightline piercing through the combined halo
of these galaxies could probe very strong \ovi\ absorption. We also  note some shared properties between extragalactic 
strong \ovi\ absorbers and \ovi\  absorption from gas related to  galactic environments:  
(i) the \ovi\ profiles generally appear featureless while the low ions
show complicated narrow absorption profiles, and (ii) absorption from both the high and 
low ionization species are observed over the entire range of velocities where
the \ovi\ absorption is observed \citep[e.g.,][]{lehner07,howk02}. 

Hence it seems possible from both observational and theoretical points of view 
that the strong \ovi\ absorber could trace coronal gas in the halo of galaxy 1339, 
especially if this galaxy is the result of a recent galaxy merger. However, this may
not be the sole explanation. In their models, \citet{oppenheimer08}  
discuss that strong \ovi\ absorbers may also be related to intragroup medium. 
The present group of galaxies must be quite different from those where $\ga 10^7$ K ($\ga 1$ keV) intragroup gas
was discovered. \citet{mulchaey96} found that X-ray detected systems contain
at least one bright elliptical galaxy ($\ga 4 L_*$) and have generally a high percentage of
early-type galaxies. With 60\% of late-type galaxies and the brightest early-type galaxy having 
$L \approx 1.6 L_*$, neither of these conditions is satisfied for the present sample of galaxies. However,
\citet{mulchaey96} also speculate that the absence of hot, diffuse intragroup medium 
in spiral-rich groups may simply mean that the hot gas was too cool to detect with {\em ROSAT},
i.e. it would have a temperature less than 0.3 keV  ($T\la 3\times 10^6$ K). This
is in fact consistent with the broadenings of the \ovi\ and \nv\ profiles. As 
the instantaneous radiative cooling is 
$t_{\rm cool} \sim (0.2$--$0.4) (n/10^{-3}\, {\rm cm}^{-3})^{-1}$ Gyr for a $10^6$ K gas \citep[e.g.,][]{gnat07}, 
if the  density is low enough, the gas can remain highly ionized for a very long time. 
Therefore, we cannot reject  a diffuse intragroup gas at $T\sim 0.1$--0.3 keV for the origin of the \ovi\ absorber. As \citet{tripp08}
show, the profiles of \ovi\ can be fitted with two components, and it is in fact quite possible
that the strong \ovi\ absorber along PKS\,0312--77 may actually trace gas from two different physical regions.

\section{The LLS and Strong \ovi\ Absorbers at Low $z$: Tracers of Circumgalactic Environments}\label{sec-bigpic}
In Table~\ref{t-lls}, we summarize the current knowledge of LLS-galaxies 
connection in the low redshift Universe. The sample of LLS observed at high spectral resolution
(allowing, e.g., to derive accurate \hi\ and metal-ion column densities) and with galaxy information is still 
small. Nevertheless this table demonstrates that LLS are not related to a single phenomenon. 
Galactic feedback, accretion of material, tidal or merger debris are all a possibility, without 
mentioning the possibility that there may be some unrelated clouds to the galaxy such as low-mass
dark matter halos (however, LLS must still be related to some galaxy activity as the 
gas is generally -- but not always -- metal enriched).  While the physical origins of the LLS may be diverse, 
LLS have also  common characteristics. First it appears obvious that the LLS are not the 
traditional interstellar gas of star-forming galaxies as DLAs could be. Second the LLS are often too metal-rich 
to be pristine intergalactic gas. Therefore, and thirdly, the LLS representing the 
IGM/galaxy interface is well supported with the current observations and knowledge.
The characteristics of the LLS are also similar to HVCs seen in the halo of the Milky Way  \citep[see, e.g.,][]{richter08} 
and found near other nearby galaxies \citep[M31, LMC, M81/M82][]{thilker04,staveley03,lehner07,chynoweth08}.
We note that the velocities of LLS relative to the host galaxy may not appear to be ``high'' (as for the LLS toward PKS\,0312--77)
on account of projection effects but also simply because in our own Galaxy, halo clouds not 
moving at high velocities cannot be separated from disk material moving at the same velocities (and indeed 
some of \hi\ clouds observed in the halo of other galaxies do not systematically show large velocity
departures from the systemic velocity of the galaxy). 

This work combined with previous studies also suggest that strong \ovi\ absorbers
generally trace circumgalactic gas rather than the WHIM, supporting the findings from 
the cosmological simulations by \citet{oppenheimer08}. We, however, emphasize that
neither in the simulations nor in the observations at low $z$ the threshold on $N($\ovi$)$ is well
defined. For example, \citet{howk08} discussed a strong \ovi\ absorber ($\log N($\ovi$) \approx 14.50$, currently 
in the top 10\% of the strongest \ovi\ absorber at $z<0.5$) 
that is more likely to be dominantly photoionized and may not be directly associated with galaxies.
However, the three strong \ovi\ absorbers discussed above are all associated with a LLS, while 
the one studied by  \citet{howk08} is not. The association of strong \ovi\ absorbers with LLS 
suggests these systems trace galactic and not intergalactic structures.

The current sample, where detailed information on the properties of the absorbers 
and galaxies in the field of view is available, is still small, but should increase in
the near future. Future observations with COS
coupled with ground based and {\em HST} imaging observations of the field of view of QSOs  will open a new door 
for studying the QSO absorbers-galaxies connection at low $z$. Attempt to systematically derive
the metallicities and SFRs of the possible host galaxies and their morphologies may help disentangling 
the various origins of the absorbers.

\section{Summary}\label{sec-sum}
We have presented multi-wavelength observations of the absorbing material at 
$z\approx 0.203$ along the QSO PKS\,0312--77 and its field of view with the goal of 
exploring the properties of the Lyman limit system and the strongest \ovi\ absorber
yet discovered in the low redshift Universe, and the connection between the absorbers 
and their environments (i.e. galaxies). The main results of our analysis are 
as follows: 

1. Using $N($\oi$)/N($\hi$)$ combined with a photoionization model, we show that the LLS 
at $z = 0.20258$ has a metallicity of about $-0.6$ dex solar. At slightly lower redshift 
($z \approx 0.2018$, velocity separation of about $-190$ \km), another absorber is 
detected with a much higher metallicity ($[{\rm Z/H}] = +0.15$) according to our 
ionization models, implying that these two absorbers have different origins.
The metallicity variation implies poor mixing of metals on galactic scale as 
observed in lower and higher redshift galactic halos. 

2. The gas in both absorbers at $z \approx 0.2018$ and $z = 0.20258$
is nearly 100\% photoionized. But only  from $-70$ to $+150$ \km\ ($0.2023 \la z \la 0.2030$), 
extremely strong \ovi\ absorption ($W_{\rm O\,VI \lambda 1032} = 493 \pm 40$ m\AA, $\log N($\ovi$) = 14.95 \pm 0.05 $)
is observed. Associated with the \ovi, there are a detection of \nv\ and a tentative 
detection of \svi.  At $z \approx 0.2018$, narrow \nv\ absorption is detected, more
consistent with the \nv\ originating from photoionized gas or in collisionally ionized gas
far from equilibrium. 

3. Using Cloudy photoionization models, we show that the high ions at $-70 \la v \la +150$ \km\
cannot be photoionized. CIE or non-equilibrium models can reproduce the observed high-ion
ratios if the gas temperature is $T\sim (3$--$10) \times 10^5$ K. The broadenings of 
\ovi\ and \nv\ are consistent with such high temperatures. The high-ion profiles are broad, 
while the low-ion profiles reveal several narrow components.
The full velocity extents of the low and high ions are, however, quite similar, as usually 
observed in galactic environments. If the gas of the 
LLS and \ovi\ absorber is cospatial, it is multiphase, with the photoionized gas
embedded within the hot, collisionally highly ionized gas. 

4. Our galaxy survey in the field of view of PKS\,0312--77 
shows that there are thirteen $0.3 \la L/L_* \la 1.6$ galaxies at $\rho \la 2 h^{-1}_{70}$ Mpc
and velocity offset from the absorbers $|\delta v| \la 1100$ \km,
implying a group of  galaxies near the \ovi\ absorber at $z \approx 0.203$. The closest 
galaxy (\#1339, a $0.7 L_*$ galaxy) has only an impact parameter of $38 h^{-1}_{70}$ kpc and is offset
by 16 \km\ from the LLS at $z = 0.20258$. There is no evidence of other $\ga 0.1 L_*$ galaxies  within 
$100 h^{-1}_{70}$ kpc. From both a visual inspection of its morphology
and its spectral classification (Sc or Irregular), galaxy 1339 appears to have resulted from a galaxy merger. 
Using diagnostics from the emission lines, we show that the metallicity of galaxy 1339 is supersolar 
($[{\rm Z/H}]_{\rm gal} = +0.15 \pm 0.05$)
and that star formation occurs at a rate  $2\pm 1$ M$_\odot$ yr$^{-1}$.  

5. Merger debris of galaxy 1339 is a very likely possibility for the origin of the LLS. Outflowing material from 
galaxy 1339 is also very probable the origin for the supersolar absorber at $z \approx 0.2018$. The 
strong \ovi\ absorber may be a tracer  of a galaxy halo fountain. 
However, the presence of a group dominated by late-type galaxies 
and with no very bright early-type galaxy may as well suggest  that the \ovi\ absorber probes 
diffuse intragroup medium at $T \la 10^6$ K. 

6. Compiling our results with other studies, it is apparent that while
the origin of the LLS is not unique (and likely includes galactic feedback, galactic halo, accreting
material,...), they must play an important role in the formation and evolution of galaxies, and are among the best 
probes to study the galaxy-IGM interface over cosmic time. Strong \ovi\ absorbers
associated with LLS appear good tracers of enrichment in galactic halos and intragroup medium
rather than the WHIM itself. 

\acknowledgments

We thank Ben Oppenheimer, Todd Tripp, and Dave Bowen for useful discussions. 
NL and JCH were supported by NASA through FUSE GI grant  NNX07AK09G and ADP grant NNX08AJ31G;
JXP acknowledges funding through an NSF CAREER grant (AST-0548180) and NSF grant (AST-0709235);
HAK was supported by NASA through grant NRA-00-01-LTSA-052. This research has made use of the NASA
Astrophysics Data System Abstract Service and the Centre de Donn\'ees de Strasbourg (CDS).


\begin{thebibliography}{}

\bibitem[Asplund, Grevesse, \& Sauval(2006)]{asplund06} 
Asplund M., Grevesse N., \& Sauval A.~J. \ 2006, CoAst, 147, 76 

\bibitem[Baldwin et al.(1991)]{baldwin91} 
Baldwin, J.~A., Ferland, G.~J., Martin, P.~G., Corbin, M.~R., Cota, S.~A., Peterson, B.~M., 
\& Slettebak, A.\ 1991, \apj, 374, 580 

\bibitem[Barton et al.(2007)]{barton07} 
Barton, E.~J., Arnold, J.~A., Zentner, A.~R., Bullock, J.~S., \& Wechsler, R.~H.\ 2007, \apj, 671, 1538

\bibitem[Bergeron \& Boiss{\'e}(1991)]{bergeron91} \
Bergeron, J., \& Boiss{\'e}, P.\ 1991, \aap, 243, 344 

\bibitem[Bertone et al.(2007)]{bertone07} 
Bertone, S., De Lucia, G., \& Thomas, P.~A.\ 2007, \mnras, 379, 1143 

\bibitem[Blanton et al.(2003)]{blanton03} 
Blanton, M.~R., et al.\ 2003, \apj, 592, 819 

\bibitem[Bouch{\'e}(2008)]{bouche08} 
Bouch{\'e}, N.\ 2008, \mnras, 389, L18 

\bibitem[Bowen et al.(2008)]{bowen08} 
Bowen, D.~V., et al.\ 2008, \apjs, 176, 59 

\bibitem[Bowen et al.(2002)]{bowen02} 
Bowen, D.~V., Pettini, M., \& Blades, J.~C.\ 2002, \apj, 580, 169 

\bibitem[Cen \& Ostriker(1999)]{cen99} 
Cen, R., \& Ostriker, J.~P.\ 1999, ApJ 514, 1

\bibitem[Cen \& Ostriker(2006)]{cen06} 
Cen, R., \& Ostriker, J.~P.\ 2006, \apj, 650, 560 

\bibitem[Chen \& Lanzetta(2003)]{chen03} 
Chen, H.-W., \& Lanzetta, K.~M.\ 2003, \apj, 597, 706 

\bibitem[Chen et al.(2001)]{chen01} 
Chen, H.-W., Lanzetta, K.~M., Webb, J.~K., \& Barcons, X.\ 2001, \apj, 559, 654 

\bibitem[Chen \& Prochaska(2000)]{chen00} 
Chen, H.-W., \& Prochaska, J.~X.\ 2000, \apjl, 543, L9

\bibitem[Churchill et al.(2005)]{churchill05} 
Churchill, C.~W., Kacprzak, G.~G., \& Steidel, C.~C.\ 2005, IAU Colloq.~199: Probing Galaxies through Quasar Absorption Lines, 24 

\bibitem[Chynoweth et al.(2008)]{chynoweth08} 
Chynoweth, K.~M., Langston, G.~I., Yun, M.~S., Lockman, F.~J., Rubin, K.~H.~R., 
\& Scoles, S.~A.\ 2008, \aj, 135, 1983 

\bibitem[Collins et al.(2003)]{collins03} 
Collins, J.~A., Shull, J.~M., \& Giroux, M.~L.\ 2003, \apj, 585, 336 

\bibitem[Cooksey et al.(2008)]{cooksey08} 
Cooksey, K.~L., Prochaska, J.~X., Chen, H.-W., Mulchaey, J.~S., 
\& Weiner, B.~J.\ 2008, \apj, 676, 262 

\bibitem[Danforth \& Shull(2008)]{danforth08} 
Danforth, C.~W., \& Shull, J.~M.\ 2008, \apj, 679, 194 

\bibitem[Dav{\' e} et al.(1999)]{dave99} 
Dav{\' e}, R., Hernquist, L., Katz, N., \& Weinberg, D.~H.\ 1999, ApJ, 511, 521 

\bibitem[de Mello et al.(2007)]{demello07}
De Mello, D.~F., Smith, L.~J., Sabbi, E., Gallagher, J.S., Mountain, M.,
\& Harbeck, D.~R. \ 2007, \aj, 135, 548

\bibitem[Dixon et al.(2007)]{dixon07} 
Dixon, W. V. et al. 2007, PASP, 119, 527

\bibitem[Ferland et al.(1998)]{ferland98} 
Ferland, G.~J., Korista, K.~T., Verner, D.~A., Ferguson, J.~W., 
Kingdon, J.~B., \& Verner, E.~M. \ 1998, PASP, 110, 761 

\bibitem[Fitzpatrick \& Spitzer(1997)]{fitzpatrick97} 
Fitzpatrick, E.~L., \& Spitzer, L.~J.\ 1997, \apj, 475, 623 

\bibitem[Gibson et al.(2000)]{gibson00} 
Gibson, B.~K., Giroux, M.~L., Penton, S.~V., Putman, M.~E., Stocke, J.~T., 
\& Shull, J.~M.\ 2000, \aj, 120, 1830 

\bibitem[Gnat \& Sternberg(2007)]{gnat07} 
Gnat, O., \& Sternberg, A.\ 2007, \apjs, 168, 213 

\bibitem[Heckman et al.(2002)]{heckman02} 
Heckman, T.~M., Norman, C.~A., Strickland, D.~K., \& Sembach, K.~R.\ 2002, \apj, 577, 691 

\bibitem[Henry, Edmunds, K{\"o}ppen(2000)]{henry00} 
Henry R.~B.~C., Edmunds M.~G., \& K{\"o}ppen J. \ 2000, \apj, 541, 660 

\bibitem[Howk et al.(2008)]{howk08} 
Howk, J.~C., Ribaudo, J., Lehner, N., Prochaska, J.~X., \& Chen, H.-W. \ 2008, \mnras, submitted 

\bibitem[Howk et al.(2002a)]{howk02a} 
Howk, J.~C., Savage, B.~D., Sembach, K.~R., \& Hoopes, C.~G.\ 2002a, \apj, 572, 264 

\bibitem[Howk et al.(2002b)]{howk02} 
Howk, J.~C., Sembach, K.~R., Savage, B.~D., Massa, D., Friedman, S.~D., 
\& Fullerton, A.~W.\ 2002b, \apj, 569, 214 

\bibitem[Hoopes et al.(2002)]{hoopes02} 
Hoopes, C.~G., Sembach, K.~R., Howk, J.~C., Savage, B.~D., 
\& Fullerton, A.~W.\ 2002, \apj, 569, 233 

\bibitem[Impey et al.(1999)]{impey99} 
Impey, C.~D., Petry, C.~E., \& Flint, K.~P.\ 1999, \apj, 524, 536 

\bibitem[Jenkins et al.(2000)]{jenkins00} 
Jenkins E.~B., et al. \ 2000, ApJ, 538, L81 

\bibitem[Jenkins et al.(2005)]{jenkins05} 
Jenkins, E.~B., Bowen, D.~V., Tripp, T.~M., \& Sembach, K.~R.\ 2005, \apj, 623, 767 

\bibitem[Jenkins et al.(2003)]{jenkins03} 
Jenkins, E.~B., Bowen, D.~V., Tripp, T.~M., Sembach, K.~R., Leighly, K.~M., Halpern, J.~P., 
\& Lauroesch, J.~T.\ 2003, \aj, 125, 2824 

\bibitem[Kennicutt(1989)]{kennicutt89} 
Kennicutt, R.~C., Jr.\ 1989, \apj, 344, 685 

\bibitem[Kennicutt(1992a)]{kennicutt92a} 
Kennicutt, R.~C., Jr.\ 1992a, \apj, 388, 310 

\bibitem[Kennicutt(1992b)]{kennicutt92b} 
Kennicutt, R.~C., Jr.\ 1992b, \apjs, 79, 255

\bibitem[Kobulnicky et al.(1999)]{kobulnicky99} 
Kobulnicky, H.~A., Kennicutt, R.~C., Jr., \& Pizagno, J.~L.\ 1999, \apj, 514, 544 

\bibitem[Kobulnicky \& Phillips(2003)]{kobulnicky03} 
Kobulnicky, H.~A., \& Phillips, A.~C.\ 2003, \apj, 599, 1031 

\bibitem[Lanzetta et al.(1995)]{lanzetta95} 
Lanzetta, K.~M., Bowen, D.~V., Tytler, D., \& Webb, J.~K.\ 1995, \apj, 442, 538 

\bibitem[Larson \& Tinsley(1978)]{larson78} 
Larson, R.~B., \& Tinsley, B.~M.\ 1978, \apj, 219, 46 

\bibitem[Lehner \& Howk(2007)]{lehner07} 
Lehner, N., \& Howk, J.~C.\ 2007, \mnras, 377, 687 

\bibitem[Lehner et al.(2008)]{lehner08} 
Lehner, N., Howk, J.~C., Keenan, F.~P., \& Smoker, J.~V.\ 2008, \apj, 678, 219 

\bibitem[Lehner et al.(2003)]{lehner03} 
Lehner N., Jenkins E.~B., Gry C., Moos H.~W., Chayer P., Lacour S. \ 2003, 
ApJ, 595, 858 

\bibitem[Lehner et al.(2007)]{lehner07a} 
Lehner, N., Savage, B.~D., Richter, P., Sembach, K.~R., Tripp, T.~M., 
\& Wakker, B.~P.\ 2007, \apj, 658, 680 

\bibitem[Lehner et al.(2006)]{lehner06} 
Lehner, N., Savage, B.~D., Wakker, B.~P., Sembach, K.~R., \& Tripp, T.~M.\ 2006, \apjs, 164, 1 

\bibitem[Lindler(2003)]{lindler03} 
Lindler, D. 2003, CALSTIS Reference Guide v7.2), (Greenbelt:NASA)

\bibitem[Lu et al.(1998)]{lu98} 
Lu, L., Sargent, W.~L.~W., Savage, B.~D., Wakker, B.~P., Sembach, K.~R., 
\& Oosterloo, T.~A.\ 1998, \aj, 115, 162 

\bibitem[Martin(1999)]{martin99} 
Martin, C.~L.\ 1999, \apj, 513, 156 

\bibitem[McGaugh(1991)]{mcgaugh91} 
McGaugh, S.~S.\ 1991, \apj, 380, 140 

\bibitem[M{\'e}nard \& Chelouche(2008)]{menard08} 
M{\'e}nard, B., \& Chelouche, D.\ 2008, \mnras, in press [arXiv:0803.0745]

\bibitem[Morton(2003)]{morton03}
Morton, D. C. \  2003, ApJS, 149, 205

\bibitem[Mulchaey et al.(1996)]{mulchaey96} 
Mulchaey, J.~S., Davis, D.~S., Mushotzky, R.~F., \& Burstein, D.\ 1996, \apj, 456, 80 

\bibitem[Osterbrock(1989)]{osterbrock89} 
Osterbrock, D.~E.\ 1989, ``Astrophysics of gaseous nebulae and active galactic nuclei", 
University Science Books 

\bibitem[Oppenheimer \& Dav{\'e}(2008a)]{oppenheimer08a} 
Oppenheimer, B.~D., \& Dav{\'e}, R.\ 2008a, \mnras, 387, 577 

\bibitem[Oppenheimer \& Dav{\'e}(2008b)]{oppenheimer08} 
Oppenheimer, B.~D., \& Dav{\'e}, R.~A.\ 2008b, MNRAS, submitted, arXiv:0806.2866 

\bibitem[Pagel et al.(1978)]{pagel78} 
Pagel, B.~E.~J., Edmunds, M.~G., Fosbury, R.~A.~E., \& Webster, B.~L.\ 1978, \mnras, 184, 569 

\bibitem[Penton et al.(2002)]{penton02} 
Penton, S.~V., Stocke, J.~T., \& Shull, J.~M.\ 2002, \apj, 565, 720 

\bibitem[Prochaska et al.(2004)]{prochaska04}
Prochaska, J.~X., Chen, H.-W., Howk, J.~C., Weiner, B.~J., 
\& Mulchaey, J.\ 2004, \apj, 617, 718 

\bibitem[Prochaska et al.(2006)]{prochaska06} 
Prochaska, J.~X., Weiner, B.~J., Chen, H.-W., \& Mulchaey, J.~S.\ 2006, \apj, 643, 680 

\bibitem[Prochter et al.(2008)]{prochter08}
Prochter, G.~E., Prochaska, J.~X., O'Meara, J.~M., Burles, S., \& Berstein, R.~A. \ 2008, \apj, submitted 

\bibitem[Proffitt et al.(2002)]{proffitt02}         
Proffitt, C., et al. 2000, STIS Instrument Handbook, v6.0, (Baltimore:STScI)

\bibitem[Rao et al.(2003)]{rao03} 
Rao, S.~M., Nestor, D.~B., Turnshek, D.~A., Lane, W.~M., Monier, E.~M., 
\& Bergeron, J.\ 2003, \apj, 595, 94 

\bibitem[Richter et al.(2008)]{richter08} 
Richter, P., Charlton, J.~C., Fangano, A.~P.~M., Bekhti, N.~B., \& Masiero, J.~R.\ 2008, \apj, submitted

\bibitem[Rauch(1998)]{rauch98} 
Rauch, M.\ 1998, \araa, 36, 267 

\bibitem[Savage et al.(1990)]{savage90} 
Savage, B.~D., Edgar, R.~J., \& Diplas, A.\ 1990, \apj, 361, 107 

\bibitem[Savage \& Sembach(1991)]{savage91} 
Savage, B. D., \& Sembach, K. R. 1991, \apj, 379, 245

\bibitem[Savage et al.(2003)]{savage03} 
Savage, B.~D., et al.\ 2003, \apjs, 146, 125 

\bibitem[Schaye et al.(2007)]{schaye07} 
Schaye, J., Carswell, R.~F., \& Kim, T.-S.\ 2007, \mnras, 379, 1169 

\bibitem[Sembach et al.(2001)]{sembach01} 
Sembach, K.~R., Howk, J.~C., Savage, B.~D., \& Shull, J.~M.\ 2001, \aj, 121, 992 

\bibitem[Sembach et al.(2003)]{sembach03} 
Sembach, K.~R., et al.\ 2003, \apjs, 146, 165 

\bibitem[Spitzer(1978)]{spitzer78} 
Spitzer, L.\ 1978, ``Physical processes in the interstellar medium", New York Wiley-Interscience

\bibitem[Steidel(1993)]{steidel93} 
Steidel, C.~C.\ 1993, Galaxy Evolution.~The Milky Way Perspective, 49, 227 

\bibitem[Staveley-Smith et al.(2003)]{staveley03} 
Staveley-Smith, L., Kim, S., Calabretta, M.~R., Haynes, R.~F., 
\& Kesteven, M.~J.\ 2003, \mnras, 339, 87 

\bibitem[Stocke et al.(2006)]{stocke06} 
Stocke, J.~T., Penton, S.~V., Danforth, C.~W., Shull, J.~M., Tumlinson, J., 
\& McLin, K.~M.\ 2006, \apj, 641, 217 

\bibitem[Thilker et al.(2004)]{thilker04} 
Thilker, D.~A., Braun, R., Walterbos, R.~A.~M., Corbelli, E., Lockman, F.~J., Murphy, E., 
\& Maddalena, R.\ 2004, \apjl, 601, L39 

\bibitem[Thom \& Chen(2008)]{thom08} 
Thom, C., \& Chen, H.-W.\ 2008, \apj, 683, 22

\bibitem[Tripp et al.(1998)]{tripp98} 
Tripp, T.~M., Lu, L., \& Savage, B.~D.\ 1998, \apj, 508, 200 

\bibitem[Tripp et al.(2008)]{tripp08} 
Tripp, T.~M., Sembach, K.~R., Bowen, D.~V., Savage, B.~D., Jenkins, E.~B., Lehner, N., 
\& Richter, P.\ 2008, \apjs, 177, 39

\bibitem[Tripp et al.(2003)]{tripp03} 
Tripp, T.~M., et al.\ 2003, \aj, 125, 3122 

\bibitem[Tumlinson \& Fang(2005)]{tumlinson05} 
Tumlinson, J., \& Fang, T.\ 2005, \apjl, 623, L97 

\bibitem[Tytler(1982)]{tytler82} 
Tytler, D.\ 1982, \nat, 298, 427 

\bibitem[Valenti et al.(2002)]{valenti02} 
Valenti, J. A., Lindler, D., Bowers, C., Busko, I., \&
Kim Quijano, J. 2002, Instrument Science Report STIS 2002-001
(Baltimore: STScI)

\bibitem[Veilleux et al.(2005)]{veilleux05} 
Veilleux, S., Cecil, G., \& Bland-Hawthorn, J.\ 2005, \araa, 43, 769 

\bibitem[Vila Costas \& Edmunds(1993)]{vila-costas93} 
Vila Costas, M.~B., \& Edmunds, M.~G.\ 1993, \mnras, 265, 199 

\bibitem[Wakker(2001)]{wakker01} 
Wakker, B.~P.\ 2001, \apjs, 136, 463 

\bibitem[Wakker et al.(2003)]{wakker03} 
Wakker, B.~P., et al.\ 2003, \apjs, 146, 1 

\bibitem[Wakker \& Savage(2008)]{wakker08} 
Wakker, B.~P., \& Savage, B.~D. \ 2008, \apjs, submitted

\bibitem[Williger et al.(2006)]{williger06} 
Williger, G.~M., Heap, S.~R., Weymann, R.~J., Dav{\'e}, R., Ellingson, E., Carswell, R.~F., Tripp, 
T.~M., \& Jenkins, E.~B.\ 2006, \apj, 636, 631 

\bibitem[Wolfe et al.(2005)]{wolfe05} 
Wolfe, A.~M., Gawiser, E., \& Prochaska, J.~X.\ 2005, \araa, 43, 861 

\bibitem[Zech et al.(2008)]{zech08} 
Zech, W.~F., Lehner, N., Howk, J.~C., Dixon, W.~V.~D., \& Brown, T.~M.\ 2008, \apj, 679, 460 

\end{thebibliography}
\end{document}